\input lanlmac
\input epsf.tex
\input mssymb.tex
\overfullrule=0pt
\def\figbox#1#2{\epsfxsize=#1\vcenter{
\epsfbox{#2}}}
\newcount\figno
\figno=0
\def\fig#1#2#3{
\par\begingroup\parindent=0pt\leftskip=1cm\rightskip=1cm\parindent=0pt
\baselineskip=11pt
\global\advance\figno by 1
\midinsert
\epsfxsize=#3
\centerline{\epsfbox{#2}}
\vskip 12pt
{\bf Fig.\ \the\figno:} #1\par
\endinsert\endgroup\par
}
\def\figlabel#1{\xdef#1{\the\figno}%
\writedef{#1\leftbracket \the\figno}%
}
\def\omit#1{}

\def\pre#1{{\tt
#1}}

\def\Rc{{\check R}}

\def\Kr{Krattenthaler}
\def\qed{\nobreak\hfill\vbox{\hrule height.4pt%
\hbox{\vrule width.4pt height3pt \kern3pt\vrule width.4pt}\hrule height.4pt}\medskip\goodbreak}

%
%
\nref\BdGN{M. T. Batchelor, J. de Gier and B. Nienhuis,
{\sl The quantum symmetric XXZ chain at $\Delta=-1/2$, alternating sign matrices and 
plane partitions},
{\it J. Phys.} A34 (2001) L265--L270,
\pre{cond-mat/0101385}.}
\nref\RSa{A. V. Razumov and Yu. G. Stroganov, {\sl Spin chains and combinatorics},
{\it J. Phys.} A34 (2001) 3185, \pre{cond-mat/0012141}\semi
{\sl Spin chains and combinatorics: twisted boundary conditions},
{\it J. Phys.} A34 (2001) 5335--5340, \pre{cond-mat/0102247}.}
\nref\RS{A. V. Razumov and Yu. G. Stroganov, 
{\sl Combinatorial nature
of ground state vector of $O(1)$ loop model},
Teor. Math. Phys.
{\bf 138} (2004) 333-337, \pre{math.CO/0104216}; 
{\sl O(1) loop model with different boundary conditions and symmetry classes 
of alternating sign matrices},
Teor. Math. Fiz. {\bf 142} (2005) 273-243, \pre{cond-mat/0108103}.}
\nref\PRdGN{P. A. Pearce, V. Rittenberg, J. de Gier and B.~Nienhuis, 
{\sl Temperley--Lieb Stochastic Processes},
J. Phys. {\bf A35} (2002) L661--L668,
\pre{math-ph/0209017}.}
\nref\dG{J.~de~Gier, {\sl Loops, matchings and alternating-sign matrices}
\pre{math.CO/0211285}.}
\nref\MNosc{S. Mitra and B. Nienhuis, {\sl 
Osculating random walks on cylinders}, in
{\it Discrete random walks}, 
DRW'03, C. Banderier and
C. Krattenthaler edrs, Discrete Mathematics and Computer Science
Proceedings AC (2003) 259-264\pre{math-ph/0312036}.} 
\nref\MNdGB{S. Mitra, B. Nienhuis, J. de Gier and M.T. Batchelor,
{\sl Exact expressions for correlations in the ground state 
of the dense $O(1)$ loop model}, J. Stat. Mech.: Theor. Exp. (2004) P09010,
\pre{cond-mat/0401245}.}
\nref\JBZ{J.-B.~Zuber, {\sl On the Counting of Fully Packed Loop 
Configurations. Some new conjectures}, Electronic J. Combin. {\bf 11(1)}
(2004) R13, \pre{math-ph/0309057}.}
\nref\DFZJZ{P.~Di~Francesco, P.~Zinn-Justin and J.-B.~Zuber,
{\sl A Bijection between classes of Fully Packed Loops and Plane Partitions},
Electronic J. Combin. {\bf 11(1)} (2004) R64, \pre{math.CO/0311220}.}
\nref\DFZ{P.~Di~Francesco and J.-B.~Zuber, {\sl On FPL 
configurations with four sets of nested arches}, 
J. Stat. Mech.: Theor. Exp. (2004) P06005, \pre{cond-mat/0403268}.}
\nref\Kratt{F. Caselli and C.~\Kr, {\sl Proof of two conjectures of
Zuber on fully packed loop configurations}, {J. Combin. Theory
Ser.} {\bf A 108} (2004) 123-146, \pre{math.CO/0312217}.  }
\nref\Krattwo{F. Caselli, C.~\Kr, B. Lass and P. Nadeau, {\sl On the number of fully packed loop
configurations with a fixed associated matching}, Electronic J. Combin. {\bf 11(2)} (2004) R16, 
\pre{math.CO/0502392}.}

\nref\PDFone{P. Di Francesco, {\sl A refined Razumov--Stroganov conjecture}, 
J. Stat. Mech.: Theor. Exp. (2004) P08009,
\pre{cond-mat/0407477}.}
\nref\PDFtwo{P. Di Francesco, {\sl A refined Razumov--Stroganov conjecture II},
J. Stat. Mech.: Theor. Exp. (2004) P11004,
\pre{cond-mat/0409576}.}
\nref\DFZJ{P.~Di Francesco and P.~Zinn-Justin, {\sl Around the Razumov--Stroganov conjecture:
proof of a multi-parameter sum rule}, \pre{math-ph/0410061}.}
\nref\IZER{A. Izergin, {\sl Partition function of the six-vertex
model in a finite volume}, Sov. Phys. Dokl. {\bf 32} (1987) 878-879.}
\nref\KOR{V. Korepin, {\sl Calculation of norms of Bethe wave functions},
Comm. Math. Phys. {\bf 86} (1982) 391-418.}
\nref\BRAU{J. De Gier and B. Nienhuis, {\sl Brauer loops and the commuting variety},
\pre{math.AG/0410392}.}
\nref\Kn{A.~Knutson, {\sl Some schemes related to the commuting variety},
\pre{math.AG/0306275}.}
\nref\IDFZJ{P.~Di Francesco and P.~Zinn-Justin, {\sl Inhomogeneous model of crossing loops
and multidegree of some algebraic varieties}, \pre{math-ph/0412031}.}
%
%
%
%
%
%
%
\nref\KZJ{A.~Knutson and P. Zinn-Justin, {\sl A scheme related to the Brauer loop model}, 
\pre{math.AG/0503224}.}
\nref\KUP{G. Kuperberg, {\sl Symmetry classes of alternating sign matrices under one roof},
Ann. of Math. {\bf 156} (2002) 835-866, \pre{math.CO/0008184}}
\nref\SKLY{E. Sklyanin, {\sl Boundary conditions for integrable quantum systems},
J. Phys. A: Mat. Gen. {\bf 21} (1988) 2375-2389.}
%
%
%
\Title{SPhT-T05/044}
{\vbox{
\centerline{Inhomogeneous loop models with open boundaries}
}}
\bigskip\bigskip
\centerline{P.~Di~Francesco,} 
\medskip
\centerline{\it  Service de Physique Th\'eorique de Saclay,}
\centerline{\it CEA/DSM/SPhT, URA 2306 du CNRS,}
\centerline{\it F-91191 Gif sur Yvette Cedex, France}
\bigskip
\vskip0.5cm
\noindent
We consider the crossing and non-crossing O(1) dense loop models on a semi-infinite
strip, with inhomogeneities (spectral parameters) that preserve the integrability.  
We compute the components of the ground state vector and obtain a closed expression
for their sum, in the form of Pfaffian and determinantal formulas.

\bigskip

\Date{04/2005}
%
%
\newsec{Introduction}

The interplay between statistical mechanics and combinatorics is an everlasting one, and takes 
many different guises as time goes. Some activity has developed recently around 
conjectural observations in Refs.\BdGN\ and \RSa-\RS\ on the ground state vectors of some simple two-dimensional
statistical models of loops, which may alternatively be viewed as one-dimensional quantum
(spin) chains. As it turned out, and among other integer numbers, the total number of alternating 
sign matrices (ASM) popped out of the study of the ground state vector of the integrable quantum
spin chain corresponding to the dense O(1) loop model on a semi-infinite cylinder of square lattice.
This number counts the total number of configurations of the ice model on a square with
domain wall boundary conditions. It
also counts the configurations of the fully packed loop model on a square grid, yet another
type of loop model, now with two kinds of loops crossing or touching at each vertex, and connecting 
by pairs the points at the periphery of the grid. 
This opened up the road to many more observations turned into conjectures, 
regarding correlation functions as well as other boundary conditions, and all involving integer sequences 
(see for instance Refs.[\xref\PRdGN-\xref\MNdGB]). An activity also developed in trying to relate
some particular subsets of configurations of the fully packed loop model to rhombus tilings of planar 
domains with possible conic singularities [\xref\JBZ-\xref\Krattwo].

The idea of considering inhomogeneous versions of the loop models came with trying to modify the boundary conditions
of the loop model on a cylinder by introducing dislocations in the underlying lattice [\xref\PDFone-\xref\PDFtwo]
and it was realized and proved in Ref.\DFZJ\ that the full multiparameter generalization of the loop model
that preserves its integrability actually leads to a ground state vector whose sum of suitably normalized
components coincides with the so-called Izergin-Korepin determinant, defined as the partition function
of the inhomogeneous six vertex model on a square grid with domain wall boundary 
conditions [\xref\IZER-\xref\KOR]. This allowed, as a by-product, to prove the conjecture of \BdGN\
that the sum of suitably normalized entries of the ground state vector of the O(1) spin chain is the
total number of alternating sign matrices. The general proof of \DFZJ\ takes full advantage of the integrability
of the model, and transforms intertwining relations for the transfer matrix of the loop model into local
recursion relations for the ground state vector's entries.

Another loop model, very similar in nature to the O(1) loop model,
also includes the possibility for loops to cross one-another. This is the so-called crossing or Brauer
O(1) loop model, for which many combinatorial conjectures were made in Ref.\BRAU, surprisingly relating 
this quantum chain to degrees of components of the commuting variety, computed in \Kn.
The same techniques as those used in \DFZJ, making full use of the integrability
of the loop model, were applied to this case in Ref.\IDFZJ, allowing to prove most of the conjectures 
of Ref.\BRAU. The algebro-geometric interpretation of these results was extended recently in Ref.\KZJ.

The two works \IDFZJ\ and \DFZJ\ are only concerned with loop models wrapped on a cylinder, i.e. with 
periodic boundary conditions. The aim of the present note is to investigate the case of open boundary
conditions, namely of (crossing or non-crossing) inhomogeneous
loop models defined on a semi-infinite strip of square
lattice. By adapting the techniques of Refs.\IDFZJ\ and \DFZJ, we will derive sum rules for the components of
the corresponding ground state vectors. The main outcome will be some particularly simple
Pfaffian or determinantal formulas for the state sum as an explicit function of the model's
inhomogeneities (spectral parameters). In the case of crossing loops, we will obtain a ``reflected"
generalization of the results of Ref.\IDFZJ, while in the non-crossing case we will be able
to identify the state sum with the partition function of so-called U-turn 
symmetric alternating sign matrices of Ref.\KUP. 
Let us stress at this point that as opposed to the crossing loop case where our proof 
is rigorous and complete, the non-crossing case relies on an assumption we make on the total degree
of the vector's components, as functions of the spectral parameters. Although we have no doubt 
this is true, proving it would certainly require a lot of effort, and we prefer to concentrate on the
consequences of this property on the ground state vector.

The paper is organized as follows. Section 2 is devoted to the case of crossing 
loops with open boundaries. After giving definitions in Sect.2.1, we show 
in Sect.2.2 the fundamental
intertwining relations satisfied by the transfer matrix of the model. Solving for all the
subsequent
relations leads to the solution of Sect.2.3 in the form of an explicit recursion relation
defining the ``fundamental" entry of the ground state vector, out of which all others
are iteratively constructed. The above relations are turned into recursion relations
for all entries of the vector in Sect.2.4, as well as into symmetry properties of the entries
in Sect.2.5. Using all these properties allows for proving two sum rules on the ground state vector
components in Sect.2.6. Section 3 is concerned with the case of non-crossing loops. We follow
the same route: definitions (Sect.3.1), intertwining properties (Sect.3.2), solution (Sect.3.3),
recursion relations (Sect.3.4), symmetries (Sect.3.5), and finally sum rule (Sect.3.6).
A few concluding remarks are gathered in Sect.4, while samples of entries
of the ground state vector are given in appendices A and B respectively for the crossing 
and non-crossing case.

\newsec{The inhomogeneous $O(1)$ crossing loop model with open boundaries}

\subsec{Transfer matrix and basic relations}

We consider the open boundary version of the inhomogeneous $O(1)$ ``Brauer" crossing loop model
considered in \IDFZJ. The latter was defined on a square lattice wrapped on a semi-infinite cylinder
of even perimeter, thus giving rise to periodic boundary conditions. We now consider the same
model on a square lattice that covers
a semi-infinite strip of width $N$ (even or odd), with centers of the lower edges labelled $1,2,...,N$. 
On each face of this domain of the square lattice, we draw at random, say with respective
probabilities $a_i,b_i,c_i$ in the $i$-th column (at the vertical of the point labelled $i$) one of the three 
following configurations
\eqn\threefaces{\epsfxsize=1.2cm \vcenter{\hbox{\epsfbox{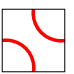} }} 
\qquad \qquad\epsfxsize=1.2cm\vcenter{\hbox{\epsfbox{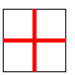}}} 
\qquad \qquad \epsfxsize=1.2cm\vcenter{\hbox{\epsfbox{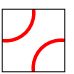}}}}
The strip is moreover supplemented with the following pattern of fixed configurations of
loops on the (left and right) boundaries:
\eqn\boundabro{\epsfxsize=6.cm\vcenter{\hbox{\epsfbox{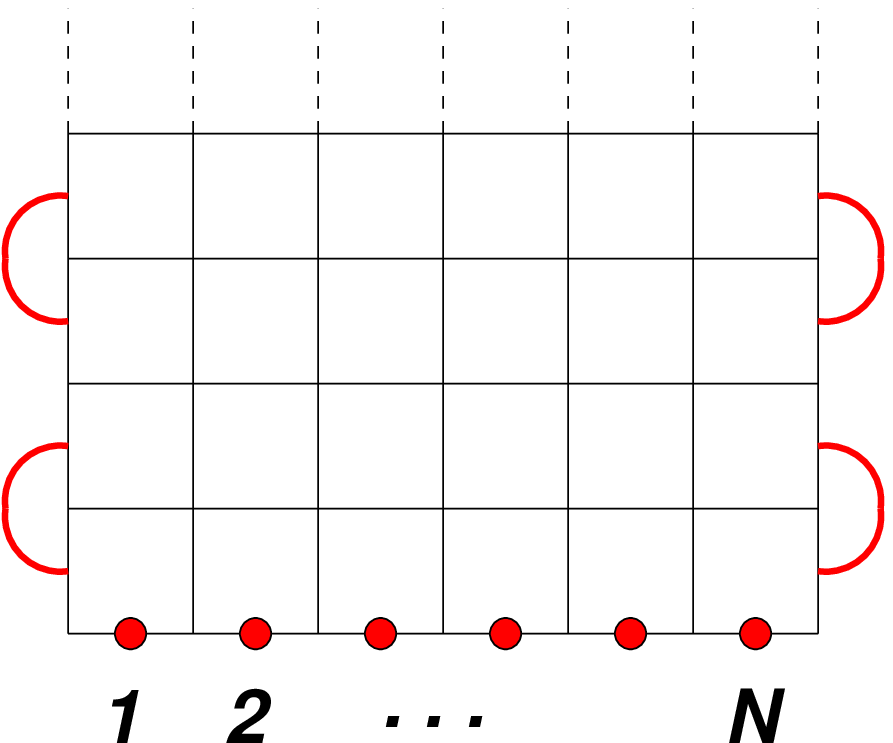}}}}
\fig{A sample configuration of the Brauer loop model on a strip
of width $N=6$ (left). We have indicated the corresponding open crossing link pattern of connection
of the points $1,2,3,4,5,6$ (right).}{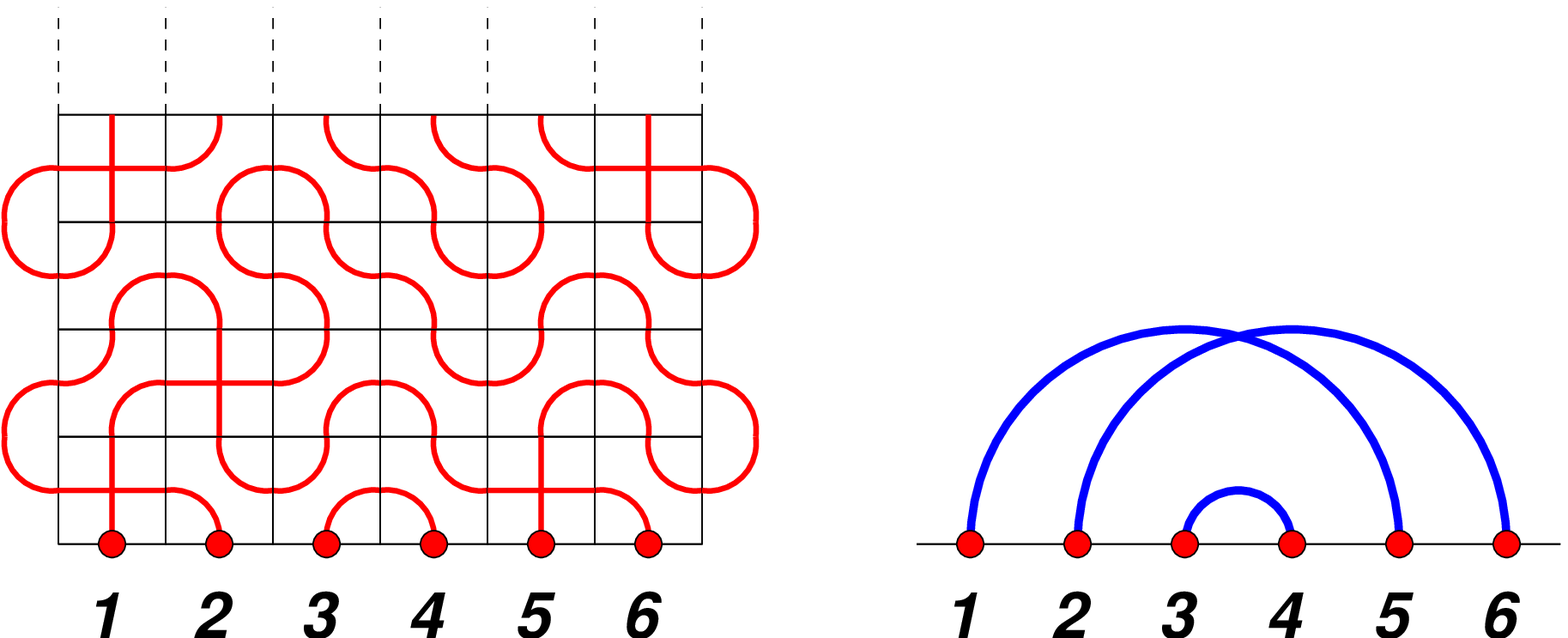}{12.cm}
\figlabel\exbrop
In a given configuration, the points $1,2,...,N$ are connected by pairs (except for one of them if
$N$ is odd, in which case it is connected to the infinity along the strip). Such a pattern
of connection is called  an {\it open crossing link pattern}. The set of open crossing link patterns 
on $N$ points is denoted by 
$CLP_{N}$, and has cardinality $(2n-1)!!$ for $N=2n$ or $N=2n-1$. 
The open crossing link patterns $\pi\in CLP_N$ span a complex vector space of dimension $(2n-1)!!$,
with canonical basis $\{\vert\pi\rangle\}_{\pi\in CLP_N}$.
An example of loop configuration
together with its link pattern are depicted in Fig.\exbrop.

One interesting question is to find
for given probability weights $a_i,b_i,c_i$, the relative probabilities of the occurrence of
the crossing link patterns $\pi\in CLP_N$. The crucial property of this loop model
is that it is integrable for the following choice of probability weights
\eqn\probintbro{ a(u)={2(1-u)\over (1+u)(2-u)}, \qquad b(u)={u(1-u)\over (1+u)(2-u)}, \qquad 
c(u)={2u\over (1+u)(2-u)}}
More precisely, the so-called $R$-matrix of the model is an operator acting on 
a vector space of open crossing link patterns or any tensor product thereof, 
say at some points labeled $i$ and $j$
of the link patterns, as follows:
\eqn\rmatbro{ R_{i,j}(z,w)= \epsfxsize=1.2cm
\vcenter{\hbox{\epsfbox{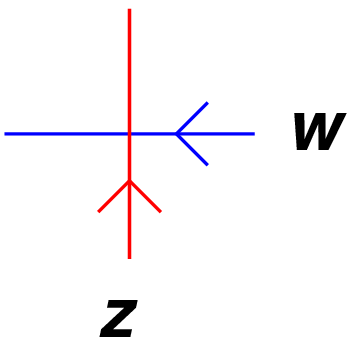}}}= a(z-w)\, \vcenter{\hbox{\epsfbox{mov1.eps}}}+ b(z-w) \,
\vcenter{\hbox{\epsfbox{mov3.eps}}} +c(z-w)\, \vcenter{\hbox{\epsfbox{mov2.eps}}}}
Here $z$ and $w$ are (arbitrary complex) spectral parameters attached respectively to the points labelled $i$ and $j$,
and are carried, as well as the point labels, by the oriented straight lines in the pictorial representation 
on the left. Each of the three possible configurations of boxes on the right acts on open crossing link patterns
as follows: the configuration must be connected by its lower
end to the point $i$ and by its right end to the point $j$ of the link patterns, thus forming new patterns
whose new points $i$ and $j$ are the upper and left ends of the box respectively.
Alternatively, the $R$-matrix 
may act locally at points $i$, $i+1$ on open crossing link patterns with $N$ points, via the permuted matrix
$\Rc=P R$:
\eqn\checkbro{\eqalign{\Rc_{i,i+1}(z,w) = \epsfxsize=1.5cm
\vcenter{\hbox{\epsfbox{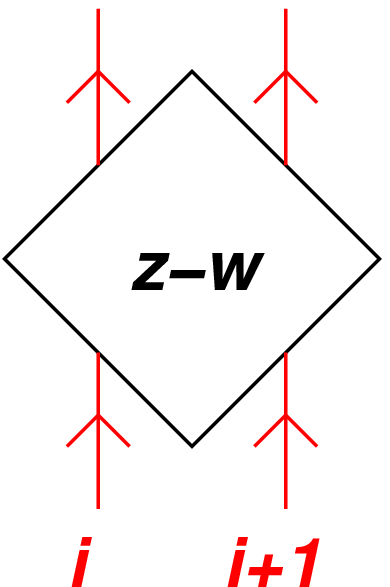}}} &= a(z-w)\, \epsfxsize=1.2cm\vcenter{\hbox{\epsfbox{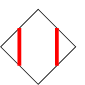} }}+ 
b(z-w)\, \vcenter{\hbox{\epsfbox{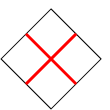} }} 
+c(z-w)\, \vcenter{\hbox{\epsfbox{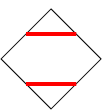} }}\cr
&=a(z-w) I\otimes I  + b(z-w) f_i  +c(z-w) e_i \cr}}
for $i=1,2,\ldots,N-1$ and 
where $P$ simply permutes the point labels, so that each label is conserved along the vertical direction.
We have displayed the matrix $\Rc$ as a linear combination of the three local operators 
$I\otimes I$, $f_i$, $e_i$, $i=1,2,...,N-1$, which form the generators of the Brauer algebra $B_N(1)$, subject
to the relations:
\eqn\bbralg{\eqalign{ e_i^2&=e_i, \qquad f_i^2=I,\qquad e_ie_{i\pm 1} e_i=e_i, \qquad
f_if_{i+1}f_i=f_{i+1}f_if_{i+1}, \cr
[e_i,e_j]&=[e_i,f_j]=[f_i,f_j]=0\ {\rm if}\ |i-j|>1, \qquad f_ie_i=e_if_i=e_i\cr}} 
These relations are clear from the pictorial representation of the action on link patterns, namely:
$I\otimes I$ leaves the link patterns unchanged,
$f_i$ crosses the links terminating at points $i$ and $i+1$, and $e_i$ glues the two ends of links at $i$
and $i+1$ and adds up a new link connecting $i$ to $i+1$. If a loop is formed in the process, it must 
simply be erased (loops are given a weight $1$ here, leading to the relation $e_i^2=e_i$).

Following Sklyanin \SKLY, we also introduce a boundary operator $K_i(z)$, whose action is diagonal at 
the points labelled $i=1$ or $N$, with matrix element 1,
but whose effect is to switch the spectral parameter $z\to -z$ attached to that point, with the 
pictorial representation
\eqn\pictobro{K_i(z)=\epsfxsize=.7cm \vcenter{\hbox{\epsfbox{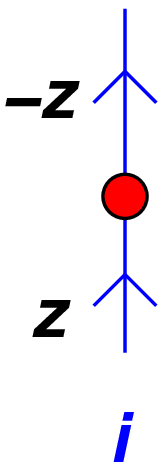} }}}

In addition to the standard Yang-Baxter and unitarity relations (with additive spectral parameters),
reading pictorially
\eqn\ybunitbro{\epsfxsize=5.cm \vcenter{\hbox{\epsfbox{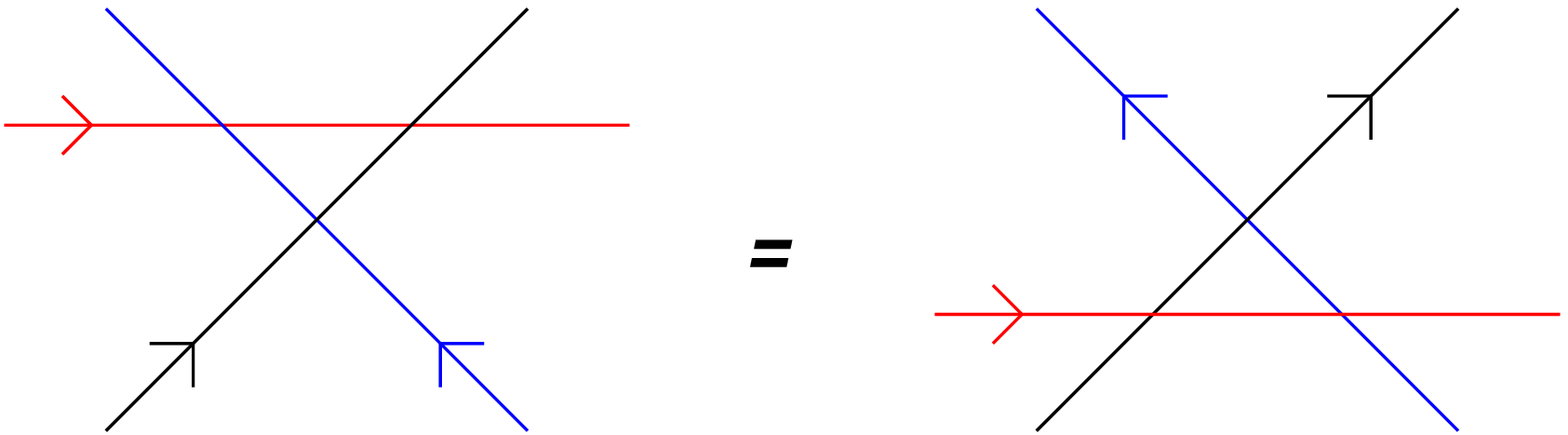} }} \quad {\rm and}\quad 
\epsfxsize=5.cm\vcenter{\hbox{\epsfbox{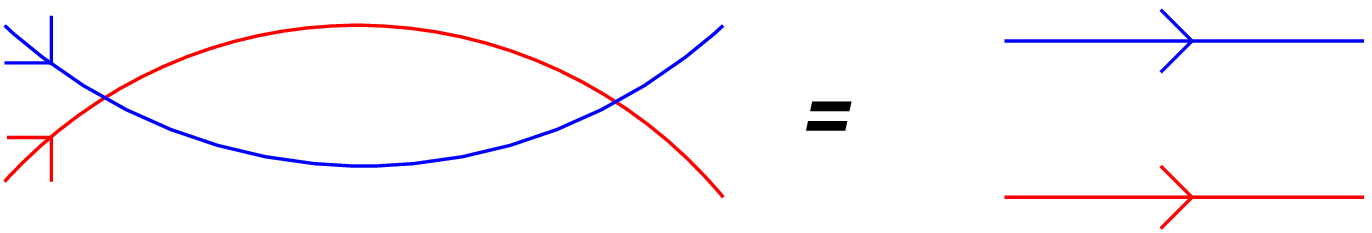} }}}
The solution to these equations for which the $\Rc$ matrix is a linear combination of generators
$I\otimes I,e_i,f_i$ is essentially unique (up to unimportant redefinitions) and takes the form
\rmatbro.
The equation \ybunitbro\ is now supplemented by the boundary Yang-Baxter relations
\eqn\bybro{\epsfxsize=11.cm \vcenter{\hbox{\epsfbox{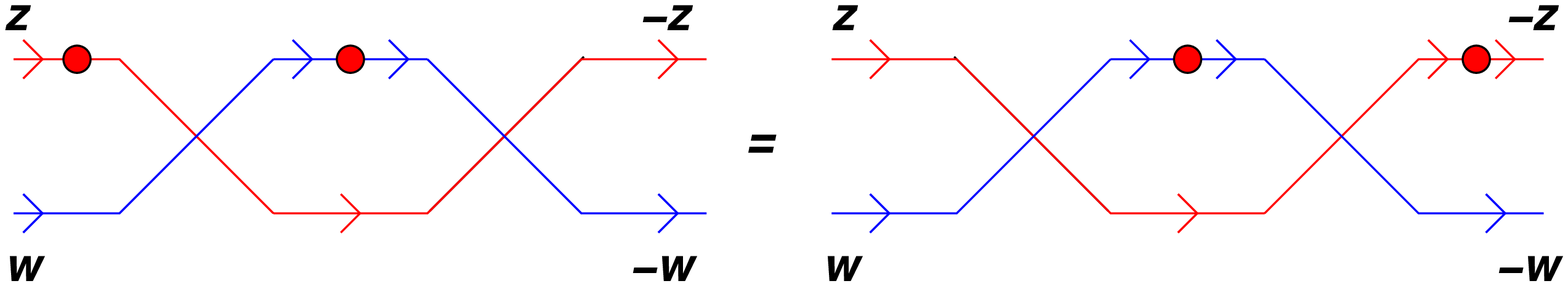} }}}
and unitarity relation
\eqn\unitbrobound{ K_i(z)K_i(-z)=I \qquad {\rm or}\ \ \epsfxsize=6.cm \vcenter{\hbox{\epsfbox{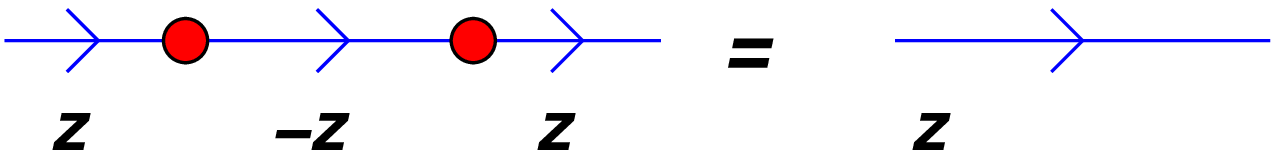} }}}
on both sides of the strip.

The transfer matrix $T(t\vert z_1,z_2,\cdots ,z_N)$ of our model reads pictorially
\eqn\pictotmbro{\epsfxsize=8.cm \vcenter{\hbox{\epsfbox{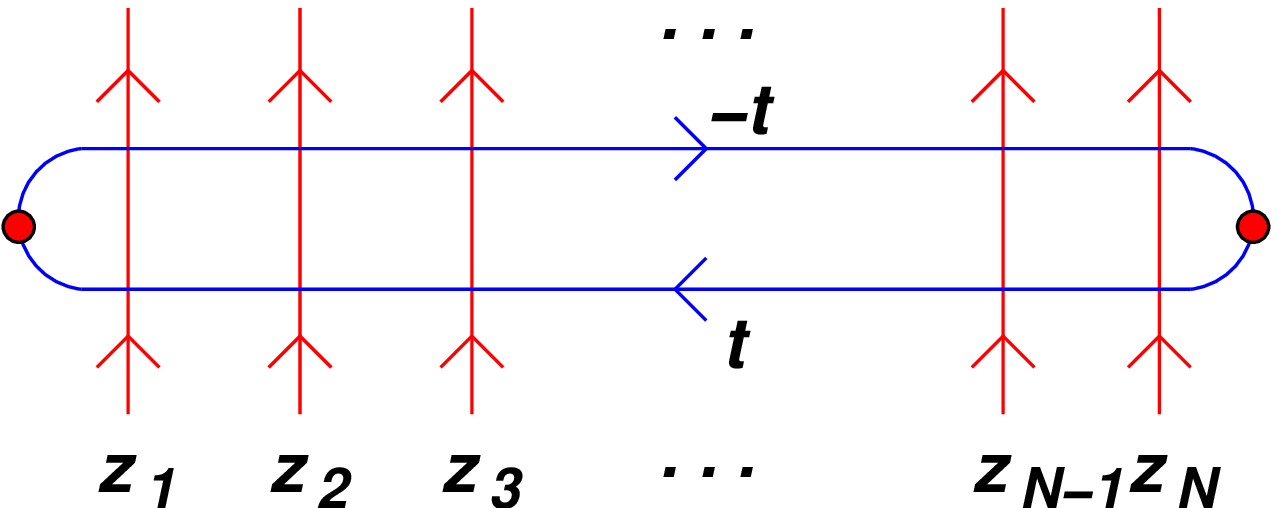} }}}
It acts from the vector space of open crossing link patterns with $N$ points to itself. 
As a consequence of the Yang-Baxter and boundary Yang-Baxter equations, the transfer matrices at two 
distinct values of $t$ commute.

We denote by $\Psi^{(N)}(z_1,z_2,\cdots ,z_N)$ the common ground state vector of the $T$'s for fixed values of
the $z_i$'s, namely such that
\eqn\eigen{ T(t\vert z_1,z_2,\cdots ,z_N) \Psi^{(N)}(z_1,z_2,\cdots ,z_N)=\Psi^{(N)}(z_1,z_2,\cdots ,z_N)}
As $T$ is a rational fraction of the $z_i$'s, we normalize $\Psi_N$ so that all its entries
are coprime polynomials of the $z_i$'s. Picking say $t=0$,
we may view the entries of $\Psi^{(N)}$ as the relative
probabilities of open link pattern connections in random crossing loop configurations 
with inhomogeneous probabilities
$(a_i,b_i,c_i)=(a(z_i),b(z_i),c(z_i))$ in each column $i$ of the strip, and the equation
\eigen expresses nothing but the invariance of probabilities under the addition of two rows 
to the semi-infinite cylinder (left and right boundaries are indeed invariant only under 
translations of {\it two} lattice spacings). This interpretation is {\it stricto sensu} only valid
in the range of $z_i$'s leading to $a_i, b_i, c_i\in [0,1]$, in which case $\Psi^{(N)}$ 
is the Perron-Frobenius eigenvector of $T$.

A last remark is in order. It turns out that the case of odd size $N=2n-1$ may always be recovered
from that of even size $N=2n$, upon taking $z_{2n}\to \infty$. Indeed, considering the transfer matrix
\pictotmbro\ of size $N$, we see that when $z_{N}\to \infty$, the two rightmost $R$-matrix elements 
(acting at the point labelled $N$) both tend to
$f_N$, hence the action at point $N$ decouples from the transfer matrix, and we have
the reduction $T(t\vert z_1,\ldots,z_N)\to T(t\vert z_1,\ldots,z_{N-1})\otimes I$, reading pictorially
\eqn\redpicto{ \epsfxsize=13.cm \vcenter{\hbox{\epsfbox{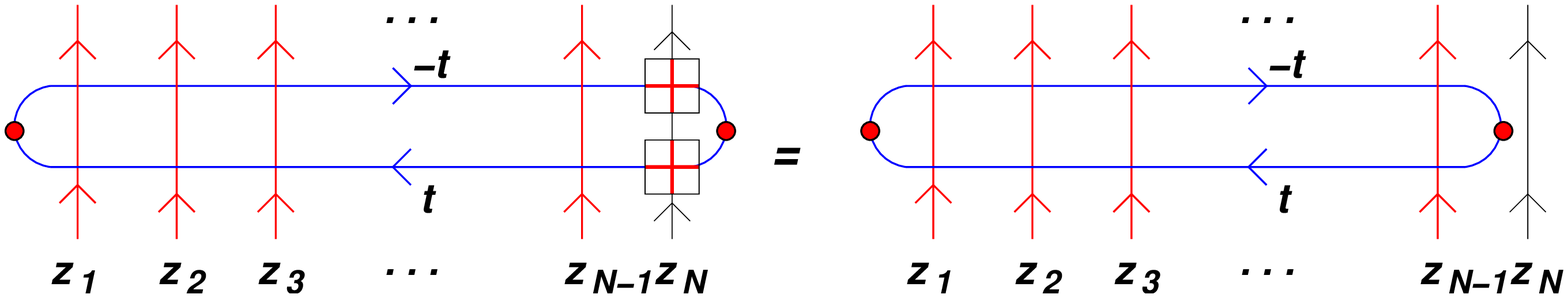}}}}
This implies that when $z_N\to\infty$, the eigenvector $\Psi^{(N)}(z_1,\ldots,z_N)$ becomes
proportional (at leading order in $z_N$) to $\Psi^{(N-1)}(z_1,\ldots,z_{N-1})$. This allows for recovering 
the odd $N$ case from the even $N$ one. Henceforth, throughout the paper
and unless otherwise specified, we will always 
assume that $N$ is even, and write $N=2n$.

\subsec{Intertwining}

As an immediate consequence of the Yang-Baxter equation, we have the intertwining property
\eqn\interl{T(t|z_1,\ldots,z_{i},z_{i+1},\ldots,z_{N})
{\check R}_{i,i+1}(z_{i},z_{i+1})
={\check R}_{i,i+1}(z_{i},z_{i+1})
T(t|z_1,\ldots,z_{i+1},z_{i},\ldots,z_{N})}
for $i=1,2,\cdots ,N-1$, also expressed pictorially as
\eqn\expict{\epsfxsize=13.cm \vcenter{\hbox{\epsfbox{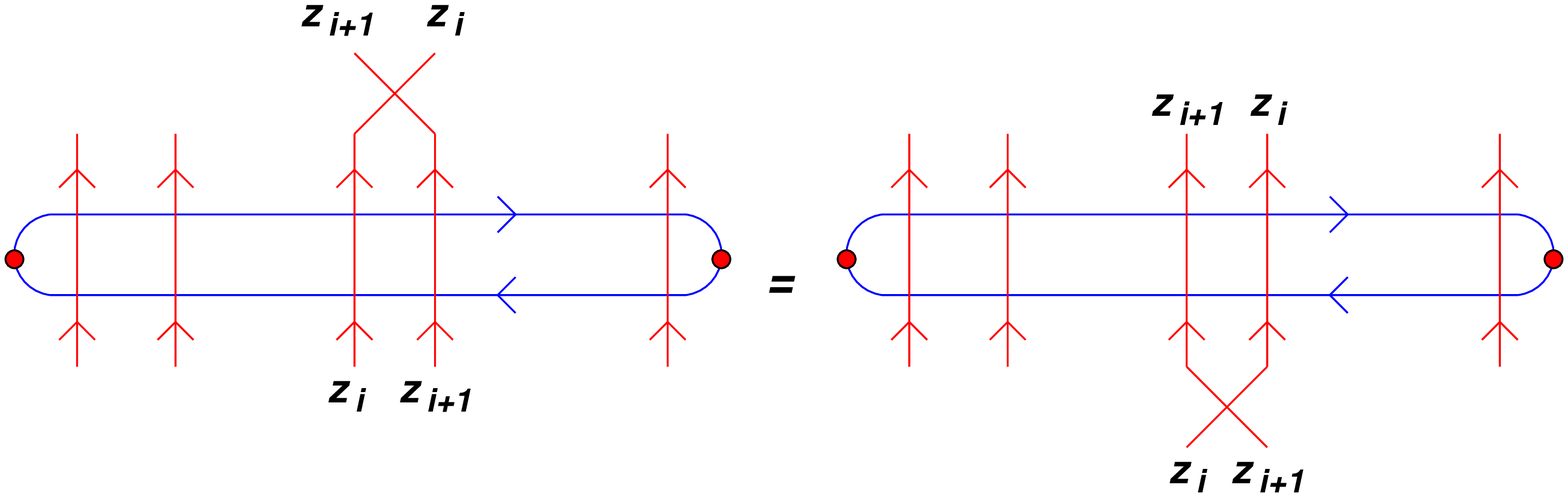} }}}
Applying this to the eigenvector $\Psi^{(N)}$ results in the relation
\eqn\transpo{\Psi^{(N)}(z_1,\ldots,z_i,z_{i+1},\ldots,z_{N})=
{\check R}_{i,i+1}(z_{i},z_{i+1})
\Psi^{(N)}(z_1,\ldots,z_{i+1},z_{i},\ldots,z_{N})}
When written in components, the latter translates
into two sets of local relations for the entries of $\Psi^{(N)}$ 
in the basis of open crossing link patterns, namely
\eqn\relabroone{
\Theta_i\Psi^{(N)}_\pi =  \Psi^{(N)}_{f_i \pi} }
for all $\pi$ with no little arch connecting points $i,i+1$
and
\eqn\relabrotwo{
\Delta_i\Psi^{(N)}_\pi= \sum_{\pi'\neq \pi \atop e_i \pi'=\pi} \Psi^{(N)}_{\pi'} }
for all $\pi$'s with a little arch joining points $i$ and $i+1$,
where $\Theta_i$ and $\Delta_i$, $i=1,2,...,N-1$, are local divided difference operators 
acting on functions of $(z_1,...,z_N)$ as
\eqn\actetdel{\eqalign{
\Theta_i&= (1+z_i-z_{i+1})\left(2\partial_i -\tau_i\right) {1\over 1+z_i-z_{i+1}} \cr
\Delta_i&= (1+z_i-z_{i+1})\left(1+{z_{i+1}-z_i\over 2}\right)\partial_i \cr}}
where $\partial_i$ and $\tau_i$ act on functions $f(z_1,...,z_N)$ as:
\eqn\actdf{\eqalign{\partial_i f &={\tau_i f- f\over z_i-z_{i+1}} \cr
\tau_i f(z_1,...,z_i,z_{i+1},...,z_N)&=f(z_1,...,z_{i+1},z_i,...,z_N)\cr}}
With these definitions, it is clear that $\Theta_i^2=I$, while $\Delta_i^2=-\Delta_i$.

An important direct consequence of Eq.\relabroone\ is that $\Psi_\pi^{(N)}$ vanishes
when $z_{i+1}=1+z_i$ if the link pattern $\pi$ has no arch joining $i$ and $i+1$.
This is easily deduced for instance from the relation \transpo\ with $R$ as in \rmatbro:
indeed, when $z_{i+1}=1+z_i$, $\Rc_{i,i+1}\propto e_i$, and therefore only components with
an arch joining $i$ to $i+1$ may be non-zero.
By taking appropriate products of $\Rc$, this was straightforwardly extended in \IDFZJ, and
$\Psi^{(N)}$ actually has the general property:
\item{\bf (P1)} For any pair $i<j$ of points such that, in the link pattern $\pi$, 
no arch connects any pair of points among $i,i+1,\ldots,j$, the component $\Psi_\pi^{(N)}$ 
vanishes when $z_j=1+z_i$.
\par

The first set of relations \relabroone\ turns out to be sufficient to generate all the entries
of $\Psi^{(N)}$ from say that corresponding to the maximally crossing link pattern, 
still denoted $\pi_0$ by a slight abuse of notation, and that connects points $i$ and $i+n$,
$i=1,2,...,n$: indeed, like in Ref.\IDFZJ, we just have to follow ``paths" from $\pi_0$ to 
$\pi=f_{i_1}f_{i_2}\cdots f_{i_k}\cdot\pi_0$ 
obtained by successive
actions of the generators $f_i$, restricted in such a way that they do not act trivially (i.e.
$f_i$ never acts on a link pattern that connects points $i$ and $i+1$) and apply \relabroone\ 
accordingly.
Any two such paths must
be equivalent modulo the braid relations $f_if_{i+1}f_i=f_{i+1}f_if_{i+1}$, $f_i^2=I$ and $f_if_j=f_jf_i$
for $|i-j|>1$. It is easy to show that the $\Theta$'s also satisfy the braid relations, just like
the ``gauged" operators
\eqn\introdelt{\delta_i=2\partial_i-\tau_i}
in terms of which $\Theta_i=(1+z_i-z_{i+1})\delta_i\ 1/(1+z_i-z_{i+1})$.
However, as already observed in Ref.\IDFZJ\ in the periodic case, the representation of the 
symmetric group they form is not faithful (it has dimension $(2n-1)!!$, to be compared with the order of
symmetric group, $(2n)!=(2n-1)!!\times 2^n n!$), and we must also implement the stabilizer relations
\eqn\stabi{\Theta_i\Theta_{i+n}\Psi_{\pi_0}^{(N)}=\Psi_{\pi_0}^{(N)}}
for $i=1,2,...,n-1$.
So, if $\Psi_{\pi_0}^{(N)}$ obeys the relations \stabi, the result of the successive actions
of $\Theta$'s yielding $\Psi_\pi^{(N)}$ out of $\Psi_{\pi_0}^{(N)}$ is independent
of the path from $\pi_0$ to $\pi$, and all components of $\Psi^{(N)}$ are therefore determined
by Eq.\relabroone\ without ambiguity.

The relations \stabi\ however
do not seem to determine $\Psi_0^{(N)}\equiv \Psi_{\pi_0}^{(N)}$ completely.  The other set of 
relations \relabrotwo\ actually serves this purpose, as we shall see in next section.
To conclude this section, let us mention two more intertwining properties, one in the ``bulk", and the other on
the boundary. The former will lead to the main recursion relation on entries of $\Psi^{(N)}$,
while the latter will allow to derive some boundary symmetry property of $\Psi^{(N)}$, both instrumental
in eventually computing the sum on the entries of $\Psi^{(N)}$.
Let us
denote by $\varphi_i$ the embedding of $CLP_{2n-2}\to CLP_{2n}$ that inserts a little arch
between points $i-1$ and $i$. We have the following restriction/projection property:
if two neighboring parameters $z_i$ and $z_{i+1}$ are
such that $z_{i+1}=1+z_1$, then
\eqn\intphibro{T(t|z_1,\ldots,z_i,z_{i+1}=1+z_i,\ldots,z_{2n})\, \varphi_i
=\varphi_i \,T(t|z_1,\ldots,z_{i-1},z_{i+2},z_{2n})}
for $i=1,2,...,N-1$. This is proved for instance in Ref.\IDFZJ\ by explicitly commuting 
$\varphi_i$ through the product of two $R$ matrices at points $i$ and $i+1$, and noting that 
when $z_{i+1}=1+z_i$, $\Rc_{i,i+1}(z_i,z_{i+1})\propto e_i$. 

Finally, using the boundary operator $K_1$ at the leftmost point, and applying the boundary 
Yang-Baxter equation \bybro, we get
\eqn\interonebro{K_1(-z_1)T(t\vert -z_1,z_2,\ldots,z_N)=T(t\vert z_1,z_2,\ldots,z_N) K_1(-z_1)}
or pictorially
\eqn\interpic{\epsfxsize=13.cm \vcenter{\hbox{\epsfbox{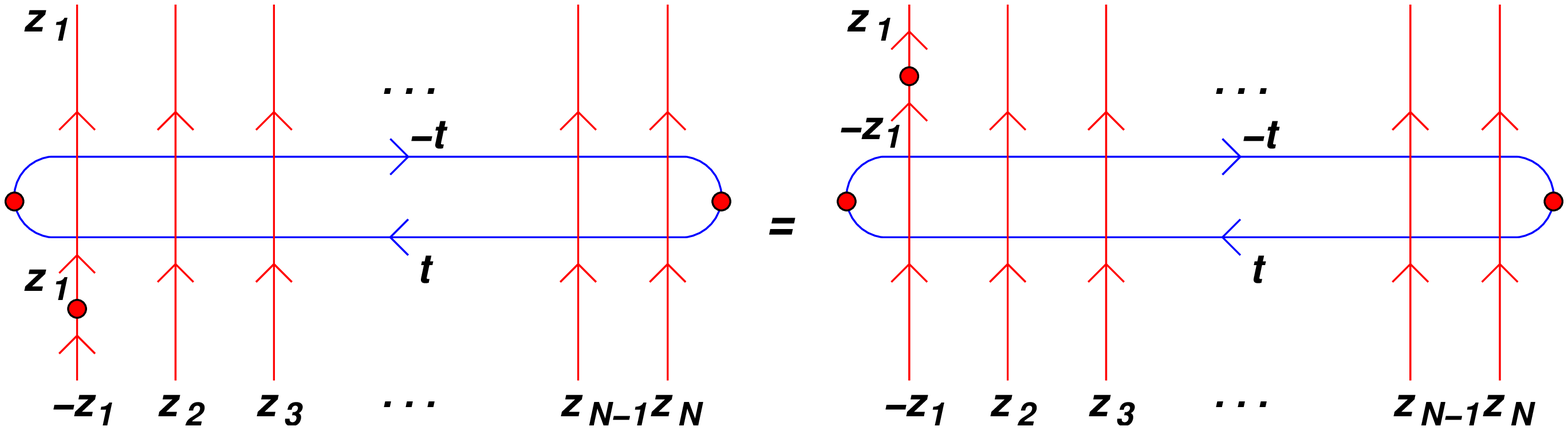}}}}
This boundary intertwining relation, when applied on the vector 
$\Psi^{(N)}(-z_1,z_2,\ldots ,z_N)$, allows to show that the latter is proportional to
$\Psi^{(N)}(z_1,z_2,\ldots ,z_N)$, and we find for even $N=2n$:
\eqn\intertwo{\Psi^{(N)}(-z_1,z_2,\ldots ,z_N)=\Psi^{(N)}(z_1,z_2,\cdots ,z_N)}
The same reasoning at the other end with the point labelled $N$ leads to the condition
\eqn\interfour{\Psi^{(N)}(z_1,z_2,\ldots z_{N-1},-z_N)=\Psi^{(N)}(z_1,z_2,\ldots ,z_N)}
In both equations, the proportionality factors are fixed to be 1 by the fact that
$\Psi^{(N)}$ is a polynomial.

Let us also mention that the system is invariant under reflection 
under which the points are reflected as $i\to N+1-i$, and the link 
patterns $\pi\to\rho(\pi)$ accordingly. Operatorwise, a global reflection reverts
all orientations of lines, and therefore inverts all operators, which amounts to
switching all $z_i\to -z_i$. As a result, the reflected eigenvector 
$\Psi^{(N)\, \rho}(-z_N,-z_{N-1},\ldots ,-z_1)$, with components
$(\Psi^{(N)\,\rho}_\pi(-z_N,-z_{N-1},\ldots ,-z_1)=\Psi^{(N)}_{\rho(\pi)}(-z_N,-z_{N-1},\ldots ,-z_1)$, 
is proportional to 
$\Psi^{(N)}(z_1,z_2,\ldots ,z_N)$, and as $\Psi^{(N)}$ is a polynomial, we have
\eqn\invervbro{\Psi^{(N)}_{\rho(\pi)}(-z_N,-z_{N-1},\ldots ,-z_1)=
\Psi^{(N)}_\pi(z_1,z_2,\ldots ,z_N)}
for all link patterns $\pi\in CLP_{N}$.

\fig{The alternative transfer matrix $T'$, together with $T$. The two commute, as a
consequence of the Yang-Baxter and boundary Yang-Baxter equations.}{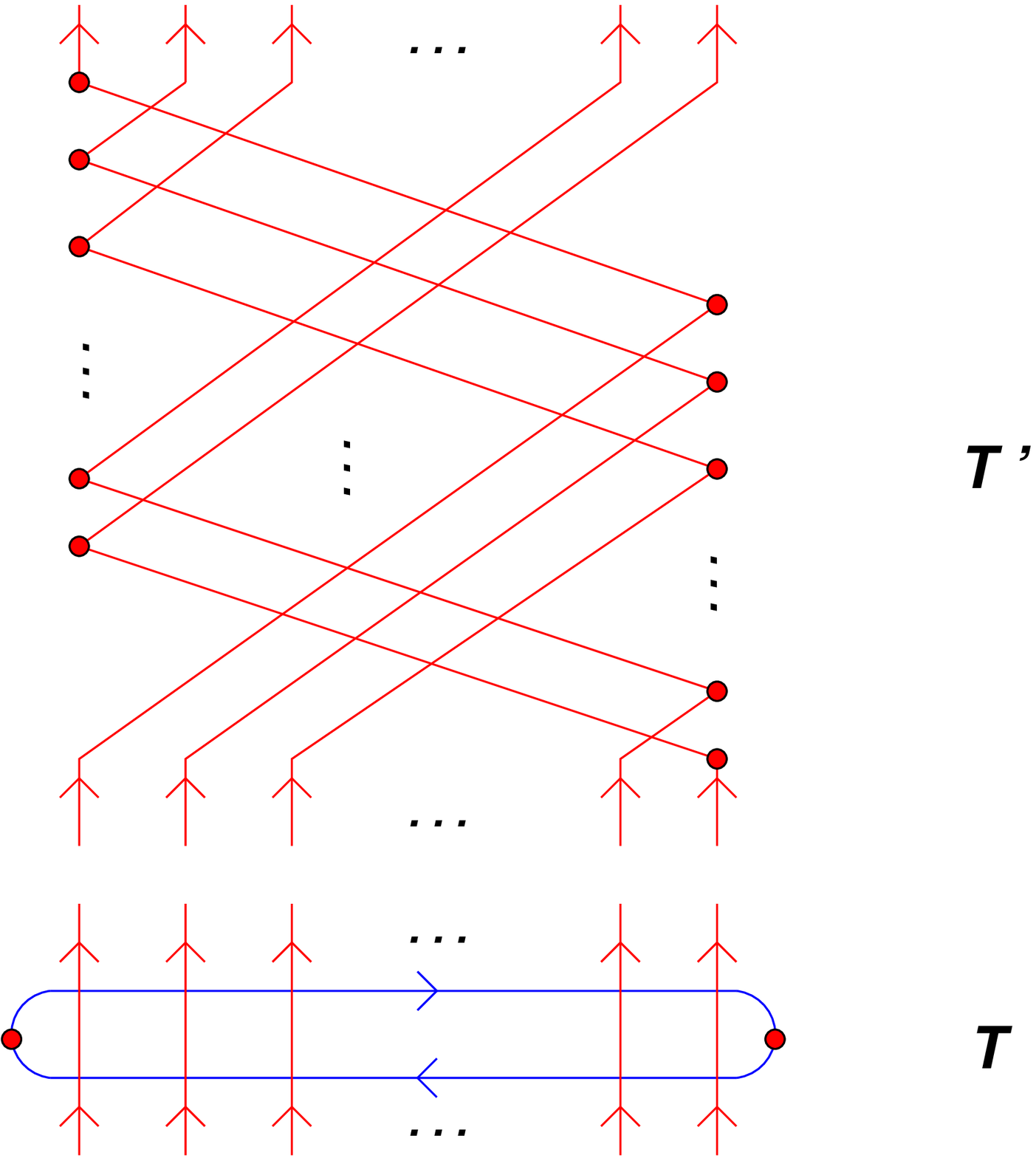}{10.cm}
\figlabel\commut

An important conclusive remark is in order. An alternative transfer matrix $T'$
for the inhomogeneous
crossing loop model with open boundary conditions may be 
written uniquely in terms of ``bulk" ($\Rc$) and ``boundary" ($K$) operators, as shown for
instance in Fig.\commut. The commutation of $T'$ with $T$ is a consequence of the Yang-Baxter
and boundary Yang-Baxter equations \ybunitbro\ and \bybro.
This means in turn that the relations on $\Psi^{(N)}$ inherited from 
the intertwining properties
involving $\Rc$ and $K$ (namely Eqs.\relabroone-\relabrotwo\ and \intertwo-\interfour)
completely determine $\Psi^{(N)}$ up to a global proportionality factor, as they produce an obvious
(Perron-Frobenius) eigenvector for $T'$. The purpose of next section
is to exhibit a candidate for $\Psi_0^{(N)}$ for which all these relations will be satisfied:
this in turn will prove, by a uniqueness argument, that the candidate for $\Psi_0^{(N)}$ is 
indeed the right value, thus solving our problem for all components of $\Psi^{(N)}$.

\subsec{Solution for $\Psi_0^{(N)}$}

As mentioned in the previous section, the relations \transpo\ together with the boundary symmetries
\intertwo-\interfour, determine $\Psi^{(N)}$ completely up to a global normalization,
which we have fixed by the coprimarity requirement, provided in addition the component
$\Psi_{\pi_0}^{(N)}=\Psi^{(N)}_{0}$ satisfies the stabilizer conditions \stabi. 
The entry $\Psi^{(N)}_{0}$ must be further determined by
the relations \relabrotwo\ and \intertwo-\interfour. The former actually reduce to just one of them, 
as we may obtain any other relation
in the list \relabrotwo\ by acting on a particular one with a succession of operators $\Theta_i$: this simply
amounts to generate any other relation by means of crossing/uncrossings of consecutive arches of the
corresponding link patterns, via an action of the $f_i$ operators that never hit little 
arches connecting $i$ to $i+1$.
Let us therefore examine only the simplest (and generic) case of 
Eq.\relabrotwo, corresponding to the link pattern $\pi=f_{n-1}f_{n-2}...f_1 \pi_0$, with a unique 
little arch, connecting points $n$ and $n+1$. In this case, Eq.\relabrotwo\ reads
\eqn\readbroto{ \Delta_n \Theta_{n-1}\Theta_{n-2}...\Theta_1\Psi^{(N)}_{0}
=(1+\Theta_n) \sum_{j=1}^{n-1} \Theta_{n-1}\Theta_{n-2}...\Theta_{j+1}
\Theta_{j-1}\Theta_{j-2}...\Theta_1 \Psi^{(N)}_{0} }
or pictorially
\eqn\bropicdel{\Delta_n \Psi\raise-.7cm\hbox{\epsfxsize=3.cm\epsfbox{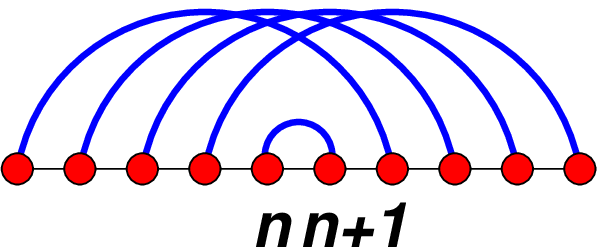}} =\sum_{j=1}^{n-1} 
\Psi\raise-.7cm\hbox{\epsfxsize=3.cm\epsfbox{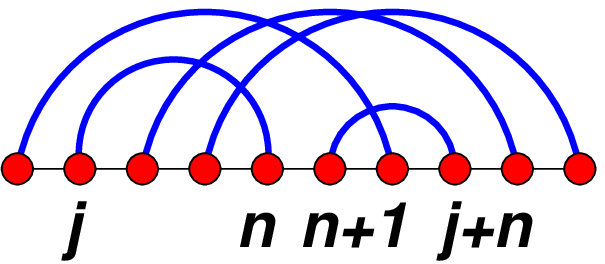}}+
\Psi\raise-.7cm\hbox{\epsfxsize=3.cm\epsfbox{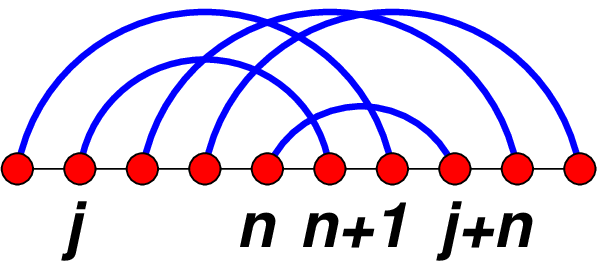}}}
Noting that $1+\Theta_i= \Delta_i \times 2/(1+z_i-z_{i+1})$, we finally get the relation
\eqn\finrelbro{ \Delta_n\left( \Theta_{n-1}\Theta_{n-2}...\Theta_1-{2\over 1+z_n-z_{n+1}}
\sum_{j=1}^{n-1} \Theta_{n-1}\Theta_{n-2}...\Theta_{j+1}
\Theta_{j-1}\Theta_{j-2}...\Theta_1\right)\Psi_{0}^{(N)} =0 }
Due to simple commutation relations between $\Theta$'s and monomials of the form $1+z_i-z_j$,
this may be recast into
\eqn\recabro{\Delta_n \ \Phi_n=0}
where 
\eqn\defi{\Phi_n=\Big(\Theta_{n-1} -{2\over 1+z_n-z_{n+1}}\Big)
\Big(\Theta_{n-2} -{2\over 1+z_{n-1}-z_{n+1}}\Big)\cdots 
\Big(\Theta_{1} -{2\over 1+z_{2}-z_{n+1}}\Big)\Psi_{0}^{(N)}}
As $\Delta_i$ is proportional to $\partial_i$, the equation \recabro\ simply expresses that
$\Phi_n$ must be invariant under the interchange of $z_n$ and $z_{n+1}$.

By explicit calculation of $\Psi^{(N)}_{0}$'s and $\Phi_n$'s
from the eigenvector condition \eigen\ for the first few values of $N=2,4,6$, we have 
observed a particularly simple formula for $\Phi_n$, which displays the desired invariance manifestly,
namely:
\eqn\formula{\eqalign{\Phi_n&=\Psi^{(N-2)}_{0}(z_1,...,z_{n-1},z_{n+2},...,z_{2n})\times \cr
&\times \prod_{i=1}^{n-1} (a_{i,n}b_{i,n}a_{i,n+1}b_{i,n+1}a_{n,n+i+1}c_{n,n+i+1}a_{n+1,n+i+1}c_{n+1,n+i+1})\cr}}
where we have defined for convenience
\eqn\definit{a_{i,j}=1+z_i-z_j, \qquad b_{i,j}=1-z_i-z_j,\qquad c_{i,j}=1+z_i+z_j}
According to previous section, the relation \formula, if true, determines $\Psi_0^{(N)}$ completely, and therefore
fixes the whole vector $\Psi^{(N)}$ as well. 
The validity of \formula\ for general $N$ also implicitly
states that no spurious overall polynomial divisor of the entries of $\Psi^{(N)}$ will occur, hence does not conflict
with the coprimarity requirement of its components.
Before using them,
let us first turn the relations \defi-\formula\ into a recursion relation for $\Psi^{(N)}_{0}$. This is readily 
done upon using the inversion formula
\eqn\invthet{ \Big( \Theta_j -{2\over a_{j+1,n+1}}\Big)^{-1}=
\Big(a_{j,n+1}\Theta_j {1\over a_{j+1,n+1}}\Big)^{-1}=
a_{j+1,n+1}\Theta_j {1\over a_{j,n+1}} =\Theta_j +{2\over a_{j,n+1}}}
where we have used $\Theta_j^2=I$. This allows to invert Eq.\formula\ into:
\eqn\recupsizerN{\eqalign{
&\Psi_0^{(N)}(z_1,...,z_{N})=\Big(\Theta_1+{2\over a_{1,n+1}}\Big)\Big(\Theta_2+{2\over a_{2,n+1}}\Big)
\times \ldots \cr 
&\ldots \times \Big(\Theta_{n-1}+{2\over a_{n-1,n+1}}\Big) 
\left( \Psi^{(N-2)}_{0}(z_1,...,z_{n-1},z_{n+2},...,z_{N})\times \right. \cr
&\left. \times \prod_{i=1}^{n-1} (a_{i,n}b_{i,n}a_{i,n+1}b_{i,n+1}a_{n,n+i+1}c_{n,n+i+1}a_{n+1,n+i+1}
c_{n+1,n+i+1})\right)\cr}}

We now state our main result: the entry $\Psi^{(N)}_0$ of the ground state vector of the crossing loop model
with open boundaries is given by the recursion relation \recupsizerN, with the initial condition that 
$\Psi_0^{(2)}=1$. To prove this statement, we must 
\item{}(i) check that $\Psi_0^{(N)}$ as given by
\recupsizerN\ is indeed a polynomial, 
\item{}(ii) check that $\Psi_0^{(N)}$ satisfies property (P1)
in order for guaranteeing the polynomiality of all other entries of $\Psi^{(N)}$,
obtained via actions of the $\Theta_i$ on $\Psi_0^{(N)}$,
and finally 
\item{}(iii) check that $\Psi^{(N)}$ thus constructed satisfies all the relations 
\relabroone-\relabrotwo, \intertwo-\interfour\ and \stabi. 
\par
\noindent Actually in the latter step, only 
\intertwo-\interfour\ and \stabi\ must be checked, as Eq.\relabroone\ is used to generate
the other entries of $\Psi^{(N)}$, and the main recursion relation also
guarantees that Eq.\relabrotwo\ is satisfied.

To check (i), we will rearrange the various factors $a,b,c$ in \recupsizerN, using 
$\Theta_{i}+{2\over a_{i,n+1}}=a_{i,i+1}a_{i+1,n+1}\delta_i \times 1/(a_{i,n+1}a_{i,n+1})$, with
$\delta_i$ as in \introdelt, and the fact that, like $\partial_i$ and $\tau_i$,
$\delta_i$ commutes with functions that are symmetric 
under the interchange $z_i\leftrightarrow z_{i+1}$.
Let us prove by induction, that for even $N=2n$:
\eqn\inducpsi{\eqalign{\Psi^{(N)}_0&(z_1,\ldots z_N)= \cr
&P^{(N)}_0(z_1,\ldots z_N)\times
\prod_{1\leq i<j\leq n}a_{i,j}b_{i,j}a_{i+n,j+n}c_{i+n,j+n}\times 
\prod_{\ell=2}^n \prod_{m=n+1}^{n+\ell-1}a_{\ell,m}\cr}}
where $P^{(N)}_0$ is a polynomial of the $z$'s. 
If we alternatively define the quantity $P_0^{(N)}$ via 
Eq.\inducpsi, we are simply left with proving that it is a polynomial.
We apply the recursion relation \recupsizerN\ to Eq.\inducpsi\ with $N\to N-2$, and with the appropriate 
shifts of variables $z_i\to z_{i+2}$ for $i=n,n+1,\ldots, 2n-2$. Commuting all possible $a,b,c$ factors
that are symmetric in $z_i,z_{i+1}$ through
the $\delta_i$'s, we finally get that Eq.\inducpsi\ is equivalent to the relation:
\eqn\obt{ P^{(N)}_0(z_1,\ldots z_N)=\prod_{i=1}^{n-1}\delta_i \, a_{i+1,n+i+1} 
\prod_{j=1}^{n-1} b_{j,n+1}c_{n,n+j+1}\, P_0^{(N-2)}(z_1,\ldots z_{n-1},z_{n+2},\ldots,z_N)}
which proves the desired result, as the operators $\delta_i$ transform polynomials into 
polynomials of the same degree: we find that $P_0^{(N)}(z_1,...,z_N)$ is a polynomial of the $z$'s,
with total degree $3n(n-1)/2$.

Moreover, Eq.\inducpsi\ allows also to immediately check the property (ii), namely that $\Psi^{(N)}_0$
has the expected vanishing properties of (P1) when $z_j=1+z_i$ (simply inspect the $a$ factors). 
Note finally that $P_0^{(2)}(z_1,z_2)=1$, as is readily seen from the explicit solution of the
eigenvector equation \eigen, and therefore using iteratively \obt, and commuting all operators
$\delta$ as much as possible to the left, we arrive at the closed expression:
\eqn\cloP{P_N(z_1,...,z_N)=\left(\prod_{r=1}^{n-1}\prod_{i=1}^{n-r}
\delta_i a_{2,n+r+i} \right)\left(\prod_{s=1}^{n-1}\prod_{j=1}^{n-s} b_{j,n+s}c_{n+1-s,n+s+j}\right)}

We are therefore left with the final task
of checking that $\Psi_0^{(N)}$, defined via \inducpsi\ and \cloP,
indeed satisfies the stabilizer property \stabi\ and
the boundary symmetry properties \intertwo-\interfour, which will in turn
be granted for any other component of $\Psi^{(N)}$, via actions with the $\Theta_i$'s. 
Eq.\stabi\ is proved in its equivalent form $\Theta_i \Psi_{0}^{(N)}=\Theta_{i+n} \Psi_{0}^{(N)}$
by induction on $n$: assume it is satisfied by $\Psi_{0}^{(N-2)}$, for $n\to n-1$. 
Going back to the original equivalent formulation 
of the recursion relation $D_n \Psi_{0}^{(N)}=\Phi_n$ with
\eqn\equirec{D_n=
\Theta_{n-1}\Theta_{n-2}\cdots \Theta_1 -{2\over a_{n,n+1}}\sum_{j=1}^{n-1}\Theta_{n-1}\Theta_{n-2}\cdots
\Theta_{j+1}\Theta_{j-1}\Theta_{j-2}\cdots\Theta_1 }
and using the braid relations satisfied by the $\Theta$'s, it is easy to show that
\eqn\commutetD{\eqalign{\Theta_i D_n&=D_n \Theta_{i+1}\cr
\Theta_{i+n+1}D_n&=D_n\Theta_{i+n+1}\cr}}
for $i=1,2,...,n-2$. Moreover, in this range of indices,
\eqn\actphitet{\eqalign{
\Theta_i \Phi_n&=\left(\prod_{i=1}^{n-1} a_{i,n}b_{i,n}a_{i,n+1}b_{i,n+1}a_{n,n+i+1}c_{n,n+i+1}a_{n+1,n+i+1}
c_{n+1,n+i+1}\right)\cr
&\times\Theta_i \Psi^{(N-2)}_{0}(z_1,...,z_{n-1},z_{n+2},...,z_{N})\cr
&=\left(\prod_{i=1}^{n-1} a_{i,n}b_{i,n}a_{i,n+1}b_{i,n+1}a_{n,n+i+1}c_{n,n+i+1}a_{n+1,n+i+1}
c_{n+1,n+i+1})\right)\cr
&\times\Theta_{i+n+1} \Psi^{(N-2)}_{0}(z_1,...,z_{n-1},z_{n+2},...,z_{N})\cr
&=\Theta_{i+n+1}\Phi_n \cr}}
where we have used the explicit symmetry of the prefactor in $z_i$ and $z_{i+1}$ and the induction hypothesis.
Combining Eqs.\commutetD\ and \actphitet, we immediately get that 
$\Theta_{i+1}\Psi_{0}^{(N)}=\Theta_{i+n+1}\Psi_{0}^{(N)}$, which amounts to Eq.\stabi\ for
$i=2,3,\ldots,n-1$ as the $\Theta$'s are involutions. 
The case $i=1$ is more tedious, as no nice commutation relations like \commutetD\ are available. However
the property $\Theta_1\Psi_{0}^{(N)}=\Theta_{n+1}\Psi_{0}^{(N)}$ reduces to
$\delta_1 a_{2,n+1}P_0^{(N)}=\delta_{n+1}a_{2,n+1}P_0^{(N)}$, which we now prove by using the explicit
expression \cloP. 

We will make extensive use of the definition \introdelt\ of $\delta_i$ and of the following
modified Leibniz rule
\eqn\modilei{ \partial_i(f g)=\tau_i(f) \partial_i(g)+g \partial_i(f)=\tau_i(g)\partial_i(f)+f\partial_i(g)}
satisfied by the divided difference operator $\partial_i$ of Eq.\actdf\ acting on the product of functions $f,g$.
When translated in terms of $\delta_i$, upon noting that $\tau_i(f g)=\tau_i(f)\tau_i(g)$, this gives:
\eqn\deltacom{ \delta_i(f g)=\tau_i(f)\delta_i(g)+2 g \partial_i(f) }
Isolating the first two terms in the product \cloP, thus writing $P_0^{(N)}=\delta_1 a_{2,n+2} R_0^{(N)}$,
we see that the condition $\delta_1 a_{2,n+1}P_0^{(N)}=\delta_{n+1}a_{2,n+1}P_0^{(N)}$
amounts to 
\eqn\amount{\eqalign{
\delta_1 a_{2,n+1}\delta_1 a_{2,n+2} R_0^{(N)}(z_1,...,z_N)
&=(a_{1,n+1}\delta_1+2)\delta_1 a_{2,n+2} R_0^{(N)}(z_1,...,z_N)\cr
&=\delta_{n+1}a_{2,n+1}\delta_1 a_{2,n+2} R_0^{(N)}(z_1,...,z_N)\cr
&=(a_{2,n+2}\delta_{n+1}+2)\delta_1 a_{2,n+2} R_0^{(N)}(z_1,...,z_N)\cr}}
where we have used the relation \deltacom\ and $\partial_1(a_{2,k})=1=\partial_k(a_{2,k})$ for $k>2$.
Eq.\amount\ amounts to $a_{1,n+1}R_0^{(N)}=\delta_{n+1}\delta_1 a_{2,n+2} R_0^{(N)}$, or equivalently
$P_0^{(N)}=\delta_1 a_{2,n+2} R_0^{(N)}=\delta_{n+1}a_{1,n+1}R_0^{(N)}$. To best illustrate the strategy 
of the proof, let us first treat the case $N=6$. First, it is easy to check directly
that  
\eqn\pfour{P_0^{(4)}(z_1,z_2,z_3,z_4)= \delta_1 a_{2,4}b_{1,3}c_{2,4}=\delta_3 a_{1,3} b_{1,3}c_{2,4}}
Then we write
\eqn\psix{\eqalign{P_0^{(6)}&= \delta_1 a_{2,5}\delta_2 a_{3,6} \delta_1 a_{2,6} Q \cr
Q&= b_{1,4}c_{3,5}b_{2,4}c_{3,6} b_{1,5}c_{2,6} \cr}}
and we have to prove that $P_0^{(6)}=S_0^{(6)}$, with
\eqn\prosix{S_0^{(6)}=\delta_4 a_{1,4} \delta_2 a_{3,6} \delta_1 a_{2,6} Q=
\delta_2 a_{3,6}\delta_4 a_{1,4}\delta_1 a_{2,6} Q}
We now wish to commute the operator $\delta_4$ all the way to the right. For this, we apply the
formula \deltacom\ to rewrite $a_{1,4}\delta_1=\delta_1a_{2,4}-2$, which yields
\eqn\yieq{\eqalign{ 
S_0^{(6)}&=(\delta_2 a_{3,6}\delta_1 a_{2,6}\delta_4 a_{2,4}-2\delta_2 a_{3,6}\delta_4 a_{2,6})Q\cr
&=(\delta_2 a_{3,6}\delta_1 a_{2,6}\delta_4 a_{2,4}-2\delta_2a_{3,6}a_{2,6}\delta_4)Q\cr}}
Now we note that 
\eqn\dfour{\eqalign{\delta_4 a_{2,4} Q&=b_{1,4}c_{3,6} b_{1,5}c_{2,6}\delta_4 a_{2,4}b_{2,4}c_{3,5}\cr
&=b_{1,4}c_{3,6} b_{1,5}c_{2,6} P_0^{(4)}(z_2,z_3,z_4,z_5)\cr
&=b_{1,4}c_{3,6} b_{1,5}c_{2,6}\delta_2 a_{3,5}b_{2,4}c_{3,5}\cr
&=\delta_2 a_{3,5} Q\cr}}
where we have used the property \pfour\ with the substitution $(z_1,z_2,z_3,z_4)\to (z_2,z_3,z_4,z_5)$.
We now take back the operator $\delta_2$ to the left:
\eqn\nyieq{\eqalign{ 
S_0^{(6)}&=(\delta_2\delta_1 a_{3,6}a_{2,6}\delta_2 a_{3,5}-2\delta_2 a_{3,6}a_{2,6}\delta_4)Q\cr
&=(\delta_2\delta_1\delta_2 a_{3,6}a_{2,6}a_{3,5}-2\delta_2 a_{3,6}a_{2,6}\delta_4)Q\cr
&=(\delta_1\delta_2\delta_1 a_{3,6}a_{2,6}a_{3,5}-2\delta_2 a_{3,6}a_{2,6}\delta_4)Q\cr
&=(\delta_1\delta_2a_{3,5}\delta_1 a_{3,6}a_{2,6}-2\delta_2 a_{3,6}a_{2,6}\delta_4)Q\cr
&=(\delta_1(a_{2,5}\delta_2+2)\delta_1 a_{3,6}a_{2,6}-2\delta_2 a_{3,6}a_{2,6}\delta_4)Q\cr
&=\big(\delta_1 a_{2,5}\delta_2 a_{3,6}\delta_1 a_{2,6}+2\delta_2 a_{3,6}a_{2,6}(\delta_2-\delta_4)
\big)Q\cr}}
where we have used the property \deltacom\ and the braid relation 
$\delta_2\delta_1\delta_2=\delta_1\delta_2\delta_1$.
Finally, we compute $(\delta_2-\delta_4)Q=b_{1,4}c_{3,6} b_{1,5}c_{2,6}(\delta_2-\delta_4)c_{3,5}b_{2,4}=0$,
again as a consequence of the property \pfour\ with the substitution $(z_1,z_2,z_3,z_4)\to (z_2,z_3,z_4,z_5)$,
which reads: $(\delta_2 a_{3,5}-\delta_4 a_{2,4})c_{3,5}b_{2,4}=0=a_{2,5}(\delta_2-\delta_4)c_{3,5}b_{2,4}$.
Subsequently, Eq.\nyieq\ reduces to $S_0^{(6)}=P_0^{(6)}$, which completes the proof for $N=6$.
We now turn to the case of general $N$. The proof is by weak induction on $N$. We assume the 
property 
\eqn\weakind{\eqalign{P_0^{(N-2r)}(z_1,...,z_{N-2r})&=\delta_1 a_{2,n-r+2}R_0^{(N-2r)}(z_1,...,z_{N-2r})\cr
&=\delta_{n-r+1} a_{1,n-r+1}R_0^{(N-2r)}(z_1,...,z_{N-2r})\cr}}
for all $r=1,2,...,n-2$.
We start from the formula for $S_0^{(N)}=\delta_{n+1}a_{1,n+1}R_0^{(N)}$:
\eqn\forS{\eqalign{S_0^{(N)}&=\delta_{n+1}a_{1,n+1} U_{2,n-1}^{(1)} U_{1,n-2}^{(2)}
U_{1,n-3}^{(3)} \ldots U_{1,2}^{(n-2)}U_{1,1}^{(n-1)} Q \cr
U_{s,t}^{(r)}&= \prod_{i=s}^{t} \delta_i a_{i+1,n+r+i} \cr
Q&=\prod_{r=1}^{n-1}\prod_{i=1}^{n-r} b_{i,n+r}c_{n+1-r,n+r+i} \cr}}
and commute the operator $\delta_{n+1}$ all the way to the right.
Actually, as $\delta_{n+1}a_{1,n+1}$ commutes with $U_{2,n-1}^{(1)}$, we simply have to commute it through
$U_{1,n-2}^{(2)}$, as $\delta_{n+1}$ then also commutes with the rest of the $U$'s on its right. 
We now use repeatedly the formula \deltacom\ to commute $\delta_{n+1}$ all the way to the right:
\eqn\rewrt{\eqalign{ \delta_{n+1}a_{1,n+1}U_{1,n-2}^{(2)}
&=\delta_{n+1}(\delta_1 a_{2,n+1}-2)a_{2,n+3}U_{2,n-2}^{(2)}\cr
&=(\delta_1 a_{2,n+3}\delta_{n+1}a_{2,n+1}-2a_{2,n+3}\delta_{n+1})U_{2,n-2}^{(2)}\cr
&=U_{1,1}^{(2)}\delta_{n+1}(\delta_2 a_{3,n+1}-2)a_{3,n+4}U_{3,n-2}^{(2)}
-2a_{2,n+3}U_{2,n-2}^{(2)}\delta_{n+1}\cr
&=U_{1,2}^{(2)}\delta_{n+1}a_{3,n+1}U_{3,n-2}^{(2)}
-2U_{1,1}^{(2)}a_{3,n+4}U_{3,n-2}^{(2)}\delta_{n+1}-2a_{2,n+3}U_{2,n-2}^{(2)}\delta_{n+1}\cr
&=U_{1,n-2}^{(2)}\delta_{n+1}a_{n-1,n+1}
-2\sum_{r=1}^{n-2} U_{1,r-1}^{(2)} a_{r+1,n+r+2}U_{r+1,n-2}^{(2)}\delta_{n+1} \cr}}
As the term $\delta_{n+1}a_{n-1,n+1}$ commutes with all $U_{1,n-r}^{(r)}$ for $r\geq 3$, we
finally get
\eqn\fingetS{\eqalign{S_0^{(N)}&=U_{2,n-1}^{(1)}U_{1,n-2}^{(2)}...U_{1,1}^{(n-1)}\delta_{n+1}a_{n-1,n+1} Q\cr
&-2\sum_{r=1}^{n-2} U_{2,n-1}^{(1)}U_{1,r-1}^{(2)} a_{r+1,n+r+2}U_{r+1,n-2}^{(2)}U_{1,n-3}^{(3)} \ldots
U_{1,1}^{(n-1)}\delta_{n+1} Q \cr}}
We now note that 
\eqn\notS{\eqalign{\delta_{n+1}a_{n-1,n+1} Q&= {Q\over b_{n-1,n+1}c_{n,n+2}} 
\delta_{n+1}a_{n-1,n+1}b_{n-1,n+1}c_{n,n+2}\cr
&= {Q\over b_{n-1,n+1}c_{n,n+2}}P_0^{(4)}(z_{n-1},z_n,z_{n+1},z_{n+2}) \cr
&= {Q\over b_{n-1,n+1}c_{n,n+2}}\delta_{n-1}a_{n,n+2}b_{n-1,n+1}c_{n,n+2}\cr}}
where we have commuted $\delta_{n+1}$ through the piece of $Q$ symmetric in $z_{n+1},z_{n+2}$
and used the property \pfour\ for $(z_1,z_2,z_3,z_4)\to (z_{n-1},z_n,z_{n+1},z_{n+2})$.
We must now take the operator $\delta_{n-1}$ to the left. Again, it is readily seen to commute with
$U_{1,n-r}^{(r)}$ for $r=n-1,n-2,...,3$. We therefore concentrate on
\eqn\concen{\eqalign{U_{2,n-1}^{(1)}&U_{1,n-2}^{(2)}\delta_{n-1}a_{n,n+2}
=U_{2,n-2}^{(1)}U_{1,n-3}^{(2)} 
\delta_{n-1}a_{n,2n} \delta_{n-2} a_{n-1,2n}\delta_{n-1}a_{n,n+2}\cr
&=U_{2,n-2}^{(1)}U_{1,n-3}^{(2)}\delta_{n-1}\delta_{n-2}\delta_{n-1}a_{n,2n}a_{n-1,2n}a_{n,n+2}\cr
&=U_{2,n-2}^{(1)}U_{1,n-3}^{(2)}\delta_{n-2}\delta_{n-1}a_{n,n+2}a_{n,2n}\delta_{n-2}a_{n-1,2n}\cr
&=U_{2,n-2}^{(1)}U_{1,n-3}^{(2)}\delta_{n-2}(a_{n-1,n+2}\delta_{n-1}+2)a_{n,2n}U_{n-2,n-2}^{(2)}\cr
&=U_{2,n-2}^{(1)}U_{1,n-3}^{(2)}\delta_{n-2}a_{n-1,n+2}\delta_{n-1}a_{n,2n}U_{n-2,n-2}^{(2)}
+2 U_{2,n-2}^{(1)}U_{1,n-3}^{(2)} a_{n,2n}a_{n-1,2n}\delta_{n-1}^2\cr
&=U_{2,n-2}^{(1)}U_{1,n-3}^{(2)}\delta_{n-2}a_{n-1,n+2}U_{n-1,n-1}^{(1)}U_{n-2,n-2}^{(2)}
+2 U_{2,n-1}^{(1)}U_{1,n-3}^{(2)}a_{n-1,2n}\delta_{n-1} \cr
&=\delta_1 a_{2,n+2}U_{2,n-1}^{(1)}U_{1,n-2}^{(2)}\cr
&+2 \sum_{r=1}^{n-2} U_{2,n-1}^{(1)}U_{1,n-r-2}^{(2)}a_{n-r,2n-r+1}\delta_{n-r}
U_{n-r+1,n-2}^{(2)}
\cr}}
where we have repeatedly used \deltacom\ and the braid relations, and also $\delta^2=1$ to
rewrite $a_{n-r+1,2n-r+1}a_{n-r,2n-r+1}=\delta_{n-r}a_{n-r+1,2n-r+1}a_{n-r,2n-r+1}\delta_{n-r}$.
Comparing Eqs.\concen\ on one hand and \rewrt\ multiplied on the left by $U_{2,n-2}^{(1)}$ on the other
hand, we find:
\eqn\wefind{\eqalign{\delta_{n+1}a_{1,n+1}U_{2,n-2}^{(1)}U_{1,n-2}^{(2)}&=
\delta_1 a_{2,n+2}U_{2,n-1}^{(1)}U_{1,n-2}^{(2)}\cr
&+2U_{2,n-1}^{(1)}
\sum_{r=1}^{n-2}U_{1,r-1}^{(2)} a_{r+1,n+r+2}(\delta_{r+1}-\delta_{n+1})U_{r+2,n-2}^{(2)}\cr}}
When multiplied by all the remaining factors $U_{1,n-3}^{(3)}...U_{1,1}^{(n-1)} Q$ on the right,
Eq.\wefind\ reads:
\eqn\finSget{S_0^{(N)}=P_0^{(N)}+2U_{2,n-1}^{(1)}\sum_{r=1}^{n-2}U_{1,r-1}^{(2)} 
a_{r+1,n+r+2}(\delta_{r+1}-\delta_{n+1})U_{r+2,n-2}^{(2)}U_{1,n-3}^{(3)}...U_{1,1}^{(n-1)} Q}
Each term in the sum is now computed by invoking the weak induction hypothesis \weakind\ for the polynomials
$P_0^{(N-2r)}(z_{r+1},z_{r+2},...,z_{N})$, which yields:
\eqn\inducweak{\eqalign{
0&=(\delta_{r+1} a_{r+2,n+2}-\delta_{n+1} a_{r+1,n+1})R_0^{(N-2r)}(z_{r+1},z_{r+2},...,z_{N})\cr
&=a_{r+1,n+1}(\delta_{r+1}-\delta_{n+1})R_0^{(N-2r)}(z_{r+1},z_{r+2},...,z_{N})\cr}}
by use of Eq.\deltacom, and where
\eqn\rdefr{ R_0^{(N-2r)}(z_{r+1},z_{r+2},...,z_{N})=U_{r+2,n-2}^{(2)}U_{r+1,n-3}^{(3)}...U_{r+1,r+1}^{(n-r-1)}Q'}
with $Q'=\prod_{s=1}^{n-r-1}\prod_{i=1}^{n-r-s} b_{i+r,n+s}c_{n+1-s,n+s+i}$. Noting finally that the
quantity
$U_{r+2,n-2}^{(2)}U_{r+1,n-3}^{(3)}...U_{r+1,r+1}^{(n-r-1)}Q'$ is exactly the piece of 
$U_{r+2,n-2}^{(2)}U_{1,n-3}^{(3)}...U_{1,1}^{(n-1)} Q$ that does not commute with 
$(\delta_{r+1}-\delta_{n+1})$, we find that each term
in the sum of \finSget\ vanishes identically. 
We conclude that $S_0^{(N)}=P_0^{(N)}$, which completes the proof that 
$\Theta_1\Psi_{0}^{(N)}=\Theta_{n+1}\Psi_{0}^{(N)}$.

To finally prove Eqs.\intertwo-\interfour,
we first note that $\Psi^{(N)}_0/P^{(N)}_0$, expressed through \inducpsi, is manifestly even in $z_1$ and $z_{2n}$,
as the quantity $a_{1,j}b_{1,j}$ is invariant under $z_1\to -z_1$, while $a_{j,2n}c_{j,2n}$
is invariant under $z_{2n}\to -z_{2n}$ for all $j$ in the range of the product.
Proceeding by induction on $N$ and 
assuming that $P^{(N-2)}_0(z_1,...,z_{N-2})$ is even in $z_1$ and $z_{N-2}$, 
we immediately get from Eq.\obt\ that $P^{(N)}_0(z_1,\ldots,z_{2n-1},-z_{2n})=P^{(N)}_0(z_1,\ldots,z_{2n-1},z_{2n})$,
as when $z_{2n}\to -z_{2n}$ the quantity $a_{n,2n}c_{n,2n}$, that carries the only dependence
on $z_{2n}$ in the prefactor, remains invariant. To prove the property for $z_1$, we simply note
that \intertwo\ is a consequence of Eqs.\interfour\ and the reflection symmetry \invervbro, which we 
just have to prove for $\Psi_{0}^{(N)}$, whose link pattern is reflection-symmetric $\pi_0=\rho(\pi_0)$.
This is again done by induction on $n$. Denoting by ${\tilde z}=(-z_{2n},-z_{2n-1},...,-z_2,-z_1)$,
we have the following property:
\eqn\propopp{ \Theta_i f(w)\bigg\vert_{w\to {\tilde z}} = \Theta_{2n-i} f({\tilde z}) }
as a direct consequence of the definition of $\Theta$. Hence performing the substitution $z\to {\tilde z}$
in $\Phi_n$ results in $\Phi_n({\tilde z})={\tilde D_n} \Psi_{0}^{(N)}({\tilde z})$, where
\eqn\tilD{ {\tilde D}_n=\Theta_{n+2}\Theta_{n+3}\cdots \Theta_{2n-1} -
{2\over a_{n,n+1}}\sum_{j=1}^{n-1}\Theta_{n+2}\Theta_{n+3}\cdots
\Theta_{n+j}\Theta_{n+j+2}\Theta_{n+j+3}\cdots\Theta_{2n-1} }
Letting this operator act on $\Psi_{0}^{(N)}(z)$ rather that on $\Psi_{0}^{(N)}({\tilde z})$, 
we may use Eqs.\stabi\ repeatedly to rewrite the result as 
${\tilde D}_n\Psi_{0}^{(N)}(z)=D_n\Psi_{0}^{(N)}(z)$: indeed, first replacing $\Theta_{2n-1}\to \Theta_{n-1}$
and commuting it all the way to the left, then repeating this for $\Theta_{2n-2}\to \Theta_{n-2}$, etc... 
until all original $\Theta$ factors are transformed, takes ${\tilde D}_n$ back to $D_n$. 
By the induction hypothesis, we have $\Phi_n({\tilde z})=\Phi_n(z)$ as all prefactors of $\Psi_{0}^{(N-2)}$
are manifestly reflection-invariant. We conclude that 
$\Phi_n({\tilde z})={\tilde D}_n\Psi_{0}^{(N)}({\tilde z})$ coincides with 
$\Phi_n(z)=D_n\Psi_{0}^{(N)}(z)={\tilde D}_n\Psi_{0}^{(N)}(z)$, hence 
$\Psi_{0}^{(N)}({\tilde z})-\Psi_{0}^{(N)}(z)$ is annihilated by the invertible operator 
${\tilde D}_n$, therefore vanishes identically, and we have proved the desired reflection symmetries.

This completes the proof that $\Psi_0^{(N)}$ is given by Eqs.\inducpsi\ and \cloP.
By inspection, as the $\Theta$'s are degree-preserving operators,
we deduce immediately that $\Psi_0^{(2n)}$ is a polynomial of total degree $4n(n-1)$ while the
partial degree in each variable is $4(n-1)$. 
In particular, in the reduction from size $N=2n$ to size $N-1=2n-1$
mentioned in Sect.2.1, we have the following formula
\eqn\limibro{ \Psi^{(N-1)}(z_1,\ldots,z_{N-1})=
\lim_{z_N\to\infty} {1\over z_N^{4(n-1)}}\Psi^{(N)}(z_1,\ldots,z_N)}

The recursion \obt\ may be implemented quite efficiently upon using the
modified Leibniz rule \modilei.
For $n=2$, we have for instance
\eqn\instfour{\eqalign{P_0^{(4)}&=(2\partial_1-\tau_1)(a_{2,4}b_{1,3}c_{2,4})\cr
&=2(b_{1,3}c_{2,4}+a_{1,4}c_{2,4}+a_{1,4}b_{2,3})-a_{1,4}b_{2,3}c_{1,4}\cr
&=b_{2,3}a_{1,4}b_{1,4}+2c_{2,4}(a_{1,4}+b_{1,3})\cr
&=5+3z_2-3z_3-2z_2z_3-z_1^2-z_4^2+(z_1^2-z_4^2)(z_2+z_3)\cr}}
where we have used the definition \actetdel, the Leibniz rule \modilei, and 
the fact that $2-c_{1,4}=b_{1,4}$. 
The explicit value of the whole vector
$\Psi^{(4)}$ in terms of the $z$'s is given in Appendix A below, 
as well as that of $\Psi_{0}^{(6)}$.
An important remark is in order.
The action of the operators $2\partial_i$ on the $a,b,c$'s
may only produce $2,-2$ or $0$ as an answer, henceforth starting from the equations \inducpsi-\obt\ 
and using the modified Leibniz formula \modilei\ repeatedly ensures by induction on $n$ that
the final result for $\Psi_{0}^{(N)}$ may be written in general as an integer linear combination
of products of $a,b,c$'s, the coefficients being only $\pm$ powers of $2$, as is the case in the third line
of Eq.\instfour. This in turn guarantees that $\Psi_{0}^{(N)}(0,0,...,0)$ is a (positive) integer in the 
homogeneous limit, and this property goes over to all other components of $\Psi^{(N)}$ via $\Theta$ actions.
To compute the values of $\Psi_{0}^{(N)}(0,0,...,0)$ using the recursion relation \obt,
we only need to know the intermediate steps $P_0^{(N-2k)}$ up to terms of degree 
$(n-k)+(n-k+1)+...+(n-1)=kn-k(k+1)/2$. With this restriction, we have found the values: 
\eqn\valpsizerobro{ 1,5,129,17369,12275137,45692809149, \ldots}
of $\Psi_{0}^{(N)}(0,0,...,0)$ for $N=2,4,6,8,10,12,\ldots$ (the value $129$ for $\Psi_{0}^{(6)}(0,0,...,0)$
may be read off Eq.(A.4) of Appendix A).

Remarkably, a recursion relation similar to \recupsizerN\ may be derived in the case of
a system with periodic boundary conditions, with even size $N=2n$. In that case, the entry 
corresponding to the maximally crossing link pattern $\pi_0$ was shown to read \IDFZJ\
\eqn\percase{\Psi_{0,{\rm per}}^{(N)}(z_1,...,z_{2n})=\prod_{i=1}^{2n}\prod_{k=1}^{n-1} a_{i,i+k} }
where the indices are taken modulo $N$ (with the convention that $i+N\equiv i$, for $i=1,2,...,N$).
Using \percase\ it is easy to prove by induction that
\eqn\recupbc{\eqalign{
&\Psi_{0,{\rm per}}^{(N)}(z_1,...,z_{2n})=\cr
&\Big(\Theta_1+{2\over 1+z_{1}-z_{n+1}}\Big)\Big(\Theta_2+{2\over 1+z_{2}-z_{n+1}}\Big) 
\cdots\Big(\Theta_{n-1}+{2\over 1+z_{n-1}-z_{n+1}}\Big) \cr
&\left(\Psi^{(N-2)}(z_1,...,z_{n-1},z_{n+2},...,z_{2n})
\prod_{i=1}^{n-1} a_{i,n}a_{i,n+1}a_{n,n+i+1}a_{n+1,n+i+1}\right) \cr}}
Actually, the ``inverse" formula analogous to \formula\ was obtained in \IDFZJ,
and was the keypoint of the proof in that case.
So, in a certain sense, the recursion relation \recupsizerN\ is a natural extension of 
the recursion relation \recupbc.

Another important property of $\Psi^{(N)}$ concerns its leading term, that is its piece of
degree $4n(n-1)$ in the $z$'s. We actually have the property:
\item{\bf (P2)} The leading terms in $\Psi_\pi^{(N)}$ read:
\eqn\leadpsibro{\Psi_\pi^{(N)}\sim (-1)^{c(\pi)}
{\prod_{1\leq i<j\leq N} (z_i^2-z_j^2)\over \prod_{{\rm pairs}\
(i<j)\atop {\rm connected}\ {\rm in}\ \pi} (z_i^2-z_j^2) } }
where $c(\pi)$ denotes the number of arch crossings in $\pi$.
\par
\noindent To prove (P2), we first show by induction on $n$ that it is satisfied by $\pi=\pi_0$.
For this, we use \obt\ and note that at large $z$'s, $\delta_i\sim -\tau_i$, as the piece $2\partial_i$
lowers the degree, and $a_{i,j}\sim (z_i-z_j)$, $c_{i,j}\sim -b_{i,j}\sim (z_i+z_j)$.
We get:
\eqn\getwe{ P^{(N)}_0 \sim 
P^{(N-2)}_0(z_1,\ldots,z_{n-1},z_{n+2},\ldots,z_N) 
\prod_{i=1}^{n-1}(z_1-z_{n+i+1})(z_{i+1}+z_{n+1})(z_{n-1}+z_{n+i+1})}
which, together with $P_0^{(2)}=1$ gives the leading behavior
\eqn\plead{ P^{(N)}_0(z_1,\ldots,z_N) \sim \prod_{i=2}^{n}\prod_{j=n+1}^{i+n-1}(z_i+z_j)
\prod_{1\leq k\leq \ell\leq n-1} (z_k^2-z_{\ell+n+1}^2) }
Once translated back in terms of $\Psi^{(N)}_0$ via \inducpsi, this yields the desired result \leadpsibro\
for $\pi=\pi_0$, with an overall sign $(-1)^{n(n-1)/2}$ from the $b$ factors, where $n(n-1)/2=c(\pi_0)$
coincides with the number of crossings in $\pi_0$. For an arbitrary link pattern
$\pi$, we simply have to act with operators $\Theta_i$ along a minimal path from $\pi_0$ to $\pi$.
As we have $\Theta_i\sim \tau_i$ for large $z$'s, Eq.\getwe\ follows immediately, as each transposition
of neighboring variables yields an overall minus sign, which parallels the fact that the number 
of crossings is decreased by 1 by the action of $f_i$ on the corresponding link pattern.

To conclude this section, let us stress that we have now given a constructive definition of $\Psi_0^{(N)}$,
leading to a vector $\Psi^{(N)}$ polynomial of total degre $4n(n-1)$ in the $z$'s. For it to match
the other definition we in Sect.2.1, we still have to check that no spurious non-trivial
polynomial factor divides all entries of $\Psi^{(N)}$ (entries are coprime). 
This will actually be proved in Sect.2.6 below,
when computing sum rules on the entries of $\Psi^{(N)}$.

\subsec{Recursion relations}

We now use the intertwining properties of Sect.2.2 to derive recursion relations
for the entries of $\Psi^{(N)}$. Given a link pattern $\pi$,
two situations may occur at a given pair of consecutive points $(i,i+1)$:

\noindent (i) the pattern $\pi$ has no arch connecting $i$ to $i+1$,
in which case property (P1) yields:
\eqn\vanish{\Psi^{(N)}_\pi(z_1,\ldots,z_{N})\Big\vert_{z_{i+1}=1+z_i}=0\ ;}

\noindent (ii) the pattern $\pi$ has a little arch joining $i$ to $i+1$, 
in which case
\eqn\recurebro{\eqalign{\Psi^{(N)}_\pi(z_1,\ldots,z_{N})\Big\vert_{z_{i+1}=1+z_i}
&=\Psi^{(N-2)}_{\pi'}(z_1,\ldots,z_{i-1},z_{i+2},\ldots,z_{N})\cr
&\times\ \prod_{\scriptstyle k=1\atop\scriptstyle k\ne i,i+1}^{N} 
(2+z_i+z_k)(2+z_i-z_k)(1+z_k-z_i)(1-z_k-z_i)\cr}}
where $\pi'$ is the link pattern $\pi$ with the little arch $i$, $i+1$
removed ($\pi=\varphi_i\pi'$, $\pi'\in CLP_{2n-2}$). 

The latter is readily obtained by applying 
Eq.\intphibro\ to the vector $\Psi^{(N)}$ at $z_{i+1}=1+z_i$, which shows proportionality between
the restricted vector $\Psi^{(N)}$ and the projected one $\Psi^{(N-2)}$, 
and the polynomial proportionality factor inbetween
is fixed by the value of $\Theta_{n-1}\Theta_{n-2}...\Theta_1\Psi^{(N)}_0$ extracted from
Eqs.\finrelbro, \defi\ and \formula.

\subsec{Symmetries}

In next section we will derive two sum rules for the components of $\Psi^{(N)}$. Before going
intop this let us display some symmetry properties of the sum over two particular sets of components
of $\Psi^{(N)}$.
We again concentrate on even $N=2n$, unless otherwise specified.

By analogy with the case of periodic boundary conditions of Ref.\IDFZJ, we may 
consider an interesting subset
of the link patterns, which we call the {\it permutation sector}, in which each link pattern only
connects points $1,2,...,n$ to points among $n+1,n+2,...,2n$. The name permutation sector
is clear, as the connections may be encoded via a permutation $\sigma\in S_n$, namely
$i \to n+\sigma(i)$ for instance. 
The simplest example of a link pattern in the permutation sector
is the maximally crossing link pattern $\pi_0$, which corresponds to the identity permutation.

Let $b_N$ denote the indicator vector of the permutation sector, with entries equal to 1
in the sector, and 0 outside. Then we have the following relations:
\eqn\realb{ b_N I=b_N,\qquad b_N f_i=b_N, \qquad b_N e_i=0}
for $i\neq n$, which lead to
\eqn\symrcb{b_N \Rc_{i,i+1}(z_i,z_{i+1})={(1-{1\over 2}(z_i-z_{i+1}))(1+z_i-z_{i+1})\over 
(1+{1\over 2}(z_i-z_{i+1}))(1-z_i+z_{i+1})} b_N
={(1+a_{i+1,i})a_{i,i+1}\over (1+a_{i,i+1})a_{i+1,i}}b_N}
Introducing 
\eqn\permuz{ W^{(N)}(z_1,...,z_N)=b_n \cdot \Psi^{(N)}(z_1,...,z_N) }
the sum over the entries of $\Psi^{(N)}$ in the permutation sector, let us act
with $b_N$ on both sides of Eq.\transpo. This immediately yields the symmetry relation
\eqn\relWtran{\eqalign{  (1+a_{i,i+1})a_{i+1,i}&W^{(N)}(z_1,\ldots,z_i,z_{i+1},\ldots,z_N)\cr
&=(1+a_{i+1,i})a_{i,i+1}W^{(N)}(z_1,\ldots,z_{i+1},z_{i},\ldots,z_N)\cr}}
valid for $i\neq n$.
More generally, replacing Eq.\transpo\ with the action of a suitable chain-product of $\Rc$'s
allows to express $\Psi^{(N)}$ as a product of consecutive $\Rc$ acting on $\Psi^{(N)}$ with $z_i$ and $z_j$
interchanged, for any pair of points $i,j$. This translates immediately 
into the generalized relation
\eqn\geneWrela{ (1+a_{i,j})a_{j,i}W^{(N)}(\ldots,z_i,\ldots,z_j,\ldots)=
(1+a_{j,i})a_{i,j}W^{(N)}(\ldots,z_j,\ldots,z_i,\ldots)} 
valid only if both $i,j\leq n$ or both $i,j>n$.
Using the boundary reflection symmetry \intertwo, we also deduce
that 
\eqn\reflW{W^{(N)}(-z_1,z_2,\ldots,z_N)=W^{(N)}(z_1,z_2,\ldots,z_N)=W^{(N)}(z_1,z_2,\ldots,-z_N)}

The same reasoning applies to the sum over all entries of $\Psi^{(N)}$. Let $v_N$ denote the vector
with all entries equal to 1, then it satisfies
\eqn\sattvN{ v_N I=v_N, \quad v_N f_i=v_N, \quad v_N e_i=v_N, \quad v_N \Rc_{i,i+1}(z_i,z_{i+1})=v_N}
Then
the sum over all components of $\Psi^{(N)}$:
\eqn\sumofallbro{ Z^{(N)}(z_1,...,z_N)=v_N \cdot \Psi^{(N)}(z_1,...,z_N) }
is symmetric in the $z$'s, as follows immediately from acting on both sides of Eq.\transpo\ with $v_N$.
$Z^{(N)}$ also satisfies the abovementioned boundary reflection symmetries under $z_1\to -z_1$ and 
$z_N\to -z_N$.

\subsec{Sum rules}
\noindent{\bf Sum rule in the permutation sector}
\par
We have for even $N=2n$ or odd $N=2n-1$:
\eqn\sumrulperbro{
W^{(N)}(z_1,...,z_N)=\prod_{1\leq i<j \leq n} a_{i,j}b_{i,j}(1+c_{i,j})(1+a_{j,i})\prod_{n+1\leq i<j\leq N}
a_{i,j}c_{i,j}(1+b_{i,j})(1+a_{j,i})}
The odd case is as a direct consequence of the even one, by application of Eq.\limibro.

To prove this for $N=2n$, let us use the symmetry relation \geneWrela: the r.h.s. of Eq.\geneWrela\ vanishes
if $z_{j}=1+z_i$ and also if $z_{j}=z_i-2$, hence $W^{(N)}(z_1,\ldots,z_N)$ must factor out 
a term $\prod_{1\leq i<j\leq n}a_{i,j}(1+a_{j,i})a_{i+n,j+n}(1+a_{j+n,i+n})$.
Note that the factors $a_{i,j}$ here correspond to the simultaneous vanishings of all the components
of $\Psi^{(N)}$ in the permutation sector, according to property (P1), as the only possible 
occurrence of a little arch connecting two consecutive points in a link pattern of the 
permutation sector is between points $n$ and $n+1$. Writing $W^{(N)}$ as
\eqn\defX{W^{(N)}(z_1,\ldots,z_N)=X^{(N)}(z_1,\ldots,z_N) 
\prod_{1\leq i<j\leq n}a_{i,j}(1+a_{j,i})a_{i+n,j+n}(1+a_{j+n,i+n}) }
for some polynomial $X^{(N)}$, we deduce from Eq.\geneWrela\ that $X^{(N)}$ is symmetric separately in
$z_1,z_2,\ldots,z_n$ and in $z_{n+1},z_{n+2},\ldots z_{2n}$. 
Moreover, Eq.\reflW\ implies that $W^{(N)}(-z_1,\ldots,z_N)=W^{(N)}(z_1,\ldots,z_N)$ has extra factors
of $\prod_{j=2}^n b_{1,j}(1+c_{1,j})$, and similarly $W^{(N)}(z_1,\ldots,-z_N)=W^{(N)}(z_1,\ldots,z_N)$
has extra factors of $\prod_{i=n+1}^{2n-1} c_{j,2n}(1+b_{j,2n})$. These two quantities
must therefore divide $X^{(N)}$, but by the abovementioned symmetries of $X^{(N)}$, it must
also be a multiple of $\prod_{1\leq i<j\leq n} b_{i,j}(1+c_{i,j})$ and of 
$\prod_{n+1\leq i<j\leq 2n} c_{i,j}(1+b_{i,j})$. This exhausts all factors in Eq.\sumrulperbro.
We have finally found a total of $4n(n-1)$ factors for $W^{(N)}$, which is therefore entirely fixed to be
given by \sumrulperbro\ up to a constant, as it is a polynomial of degree $4n(n-1)$.
The constant is now further fixed to be 1 by the leading term of $W^{(N)}$. Indeed, using the property (P2)
Eq.\leadpsibro, we may write the leading term in $W^{(N)}$ as
\eqn\ledW{\eqalign{W^{(N)}&\sim \Delta(z_1^2,\ldots,z_{2n}^2) \sum_{\sigma\in S_n} {{\rm sgn}(\sigma) \over
\prod_{i=1}^n (z_i^2-z_{n+\sigma(i)}^2) } \cr
&=\Delta(z_1^2,\ldots,z_{2n}^2) \det\Big( {1\over (z_i^2-z_{n+i}^2})\Big)_{1\leq i,j\leq n} \cr
&=\Delta(z_1^2,\ldots,z_{2n}^2){\Delta(z_1^2,\ldots,z_n^2)\Delta(z_{n+1}^2,\ldots,z_{2n}^2)\over
\prod_{i,j=1}^n z_i^2-z_{j+n}^2}\cr
&=\Delta(z_1^2,\ldots,z_n^2)^2\Delta(z_{n+1}^2,\ldots,z_{2n}^2)^2
\cr}}
where we have used the shorthand notation $\Delta(m_1,...,m_p)$ for the Vandermonde determinant
$\prod_{1\leq i<j\leq p}(m_i-m_j)$, applied the parametrization of link patterns in the
permutation sector by the permutations $\sigma\in S_n$,
interpreted $(-1)$ to the number of crossings as the signature of the permutation, 
and finally applied the
Cauchy determinant formula to reexpress the resulting determinant as a product. 
The leading behavior of the r.h.s. of 
Eq.\sumrulperbro\ is readily checked to coincide with this product.

As a side result of the sum rule \sumrulperbro,
we conclude that the vector $\Psi^{(N)}$ constructed in Sect.2.3 indeed satisfies the coprimarity
constraint on its components, as its degree must be at least $4n(n-1)$ from
the necessary factors of its sum rule within the permutation sector, leaving no place 
for overall spurious polynomial factors.

Note finally that in the homogeneous limit where all $z\to 0$, we simply get the integers
\eqn\homoW{\eqalign{ W^{(2n)}(0,0,\ldots,0)&=2^{2n(n-1)} \cr
W^{(2n-1)}(0,0,\ldots,0)&=2^{2(n-1)^2} \cr}}
as the sum of the integer entries of $\Psi^{(N)}(0,\ldots,0)$ in the permutation sector.

\noindent{\bf Sum rule for all components of $\Psi^{(N)}$}
\par
We have for even $N=2n$:
\eqn\sumrulbrau{\eqalign{Z^{(N)}(z_1,..,z_{N})&=\prod_{1\leq i<j\leq N}  {1-(z_i-z_j)^2\over z_i-z_j}
{1-(z_i+z_j)^2\over z_i+z_j} \cr
&\times {\rm Pf}
\left({z_i-z_j\over 1-(z_i-z_j)^2}{z_i+z_j\over 1-(z_i+z_j)^2}\right)_{1\leq i<j\leq N}\cr}}
while for odd $N=2n-1$, we have
\eqn\oddsumrulbrau{Z^{(N-1)}(z_1,..,z_{N-1})=\lim_{z_{N}\to \infty} {1\over z_{N}^{4(n-1)}} Z^{(N)}(z_1,..,z_{N})}
The latter relation is a consequence of Eq. \limibro.

The relation \sumrulbrau\ is proved by induction on $n$. As mentioned in Sect.2.5, 
$Z^{(N)}$ is a symmetric polynomial of the $z$'s, and by a similar reasoning as above, we also conclude that
$Z^{(N)}(z_1,...,z_{i-1},-z_i,z_{i+1},...,z_N)=Z^{(N)}(z_1,...,z_{i-1},z_i,z_{i+1},...,z_N)$, for all $i=1,2,...,N$.
Moreover, from the properties (i-ii) of Sect.2.4, and upon summing over the entries
of $\Psi^{(N)}$, we see that $Z^{(N)}$ satisfies for instance the recursion relation
\eqn\recusatZbro{\eqalign{ Z^{(N)}(z_1=z_2-1,z_2,&z_3,...,z_N)= Z^{(N-2)}(z_3,z_4,...,z_N)\cr
&\times \ \prod_{k=3}^N (1+z_2+z_k)(1+z_2-z_k)(2+z_k-z_2)(2-z_k-z_2) \cr}}
By the above symmetries, this fixes the value of $Z^{(N)}$ for $z_1=\pm(\pm z_k-1)$, $k=2,3,...,N$,
hence a total of $4(n-1)$ values,
which fixes $Z^{(N)}$ as a function of $z_1$ up to a proportionality constant, as $Z^{(N)}$ is a polynomial 
of degree $4(n-1)$ of $z_1$. The latter is further fixed by writing the leading behavior for large
$z$'s of $Z^{(N)}$:
\eqn\larNzN{\eqalign{Z^{(N)}&\sim \Delta(z_1^2,\ldots,z_{2n}^2)\ \sum_{\pi\in CLP_{2n}} {(-1)^{c(\pi)}\over
\prod_{{\rm pairs}\ (i<j)\atop {\rm connected}\ {\rm by}\ \pi} (z_i^2-z_j^2)}\cr
&=\Delta(z_1^2,\ldots,z_{2n}^2) \ {\rm Pf}\left( {1\over (z_i^2-z_j^2)}\right)_{1\leq i<j\leq 2n}\cr}}
where we have directly identified the Pfaffian, upon interpreting the $\pi$'s as permutations of $S_{2n}$
with only cycles of length $2$, and $(-1)^{c(\pi)}$ as the signature of the corresponding permutation.
The r.h.s. of \sumrulbrau\ is clearly a polynomial of the $z$'s, symmetric under interchange and sign reversal
of the $z$'s,  of total degree $4n(n-1)$ and partial degree $4(n-1)$ in each variable. 
Moreover, it clearly satisfies the recursion relation \recusatZbro, as when $z_1\to z_2-1$ the first two
lines and columns of the matrix $A_{i,j}=(z_i^2-z_j^2)/((1-(z_i-z_j)^2)(1-(z_i+z_j)^2)$ are
dominated by the terms $A_{1,2}=-A_{2,1}$, henceforth the determinant of $A$ factors into that of $A$
with the first two rows and columns deleted, and the proportionality factor coming from
the prefactor in \sumrulbrau\ matches that in \recusatZbro. Moreover, for large $z$'s, the 
Pfaffian reduces to
\eqn\gigapfaff{ \prod_{1\leq i<j\leq N}(z_i^2-z_j^2)\ 
{\rm Pf}\left({1\over z_i^2-z_j^2}\right)_{1\leq i<j\leq N}}
and matches exactly the sum over the leading terms of the components of $\Psi^{(N)}$, as given by 
Eq.\larNzN. This completes the proof of the sum rule \sumrulbrau.

Note finally that in the homogeneous limit where all $z\to 0$, the formula \sumrulbrau\ reduces to
\eqn\homoZbro{ Z^{(2n)}(0,0,\ldots,0)= 
{\rm Pf}\left( {1\over 2}\Big({2i+2j+1\choose 2j}-{2i+2j+1\choose 2i}\Big)\right)_{0\leq i<j\leq 2n-1}}
while in the odd $N=2n-1$ case, we have
\eqn\oddhomobro{ Z^{(2n-1)}(0,0,\ldots,0)={\rm Pf}\left( {1\over 2}\Big({2i+2j+1\choose 2j}-
{2i+2j+1\choose 2i}\Big)\right)_{1\leq i<j\leq 2n-2}}
The numbers $Z^{(N)}(0,\ldots,0)$ read
\eqn\numzzerbro{1, 7, 39, 1771, 57163, 16457953, 3125503009, 5643044005273,6357601085989209,\ldots}
for $N=2,3,4,\ldots,10,\ldots$

\newsec{The inhomogeneous $O(1)$ loop model with open boundaries} 

We now turn to the open boundary version of the inhomogeneous $O(1)$ (non-crossing) loop model
considered in Refs.[\xref\BdGN-\xref\DFZJ].
Throughout this section and Appendix B, we use the same notations for transfer matrices,
ground state vectors, fundamental link patterns, etc... as in the Brauer case, as there is no ambiguity
that from now on we change the subject and deal with a different case. This allows for avoiding many repetitions,
as many of the equations of the Brauer case still hold in the non-crossing one.

\subsec{Transfer matrix and basic relations}

Like in the Brauer case,
the model of non-crossing loops was originally defined 
on a square lattice wrapped on a semi-infinite cylinder
of even perimeter, giving rise to periodic boundary conditions. 
We now consider the same
model on a square lattice that covers
a semi-infinite strip of width $N$ (even or odd), with centers of the lower edges labelled $1,2,...,N$.
We attach probabilities $(t_i,1-t_i)$ to the two face loop configurations 
\eqn\twoconfigs{\epsfxsize=1.2cm\vcenter{\hbox{\epsfbox{mov1.eps}}} \quad {\rm and} 
\quad \epsfxsize=1.2cm\vcenter{\hbox{\epsfbox{mov2.eps}}}}
in the column above the edge labelled $i$. We moreover supplement the picture with the {\it same} 
patterns of fixed configurations of loops on the (left and right) boundaries as for the Brauer case, 
as depicted in Eq.\boundabro. A given configuration now forms a planar pairing of the $N$ labelled points
(one of which is connected to infinity if $N$ is odd), via $n=[N/2]$ non-crossing arches. 
The set of such link patterns is denoted by $LP_N$,
and has cardinality $c_n={2n\choose n}/(n+1)$ the $n$-th Catalan number, for $N=2n$ or $N=2n-1$.
As before, we also consider the $c_n$-dimensional complex vector space with canonical basis indexed by elements
of $LP_N$.

The transfer matrix
for this system is built out of the basic $R$-matrix, that acts on open link patterns or tensor products thereof
via:
\eqn\rmat{ R_{i,j}(z,w)= \epsfxsize=1.2cm
\vcenter{\hbox{\epsfbox{rmat.eps}}}= t(z,w)\, \vcenter{\hbox{\epsfbox{mov1.eps}}}+\big(1-t(z,w)\big)
\, \vcenter{\hbox{\epsfbox{mov2.eps}}} }
where as before $z$ and $w$ are spectral parameters attached to the points labelled $i$ and $j$ respectively,
and we use the same pictorial representation for the matrix elements of $R$ (intersection between two
oriented lines carrying the spectral parameters $z$ and $w$). Alternatively, we have the permuted
matrices
\eqn\checR{\eqalign{\Rc_{i,i+1}(z,w)&= t(z,w)\, \epsfxsize=1.2cm\vcenter{\hbox{\epsfbox{id.eps} }}+ 
+\big(1-t(z,w)\big)\, \vcenter{\hbox{\epsfbox{ei.eps} }}\cr
&=t(z,w) I\otimes I  +\big(1-t(z,w)\big)\, e_i \cr}}
for $i=1,2,\ldots,N-1$
acting on the vector space of link patterns,
where $e_i$, $i=1,...,N-1$ are now the generators of the Temperley-Lieb algebra $TL_N(1)$, subject
to the relations
\eqn\tla{ e_i^2=e_i,\qquad [e_i,e_j]=0\ {\rm if}\ |i-j|>1,\qquad e_ie_{i\pm 1}e_i=e_i}
While $I\otimes I$ leaves link patterns invariant, $e_i$ glues the ends labelled $i$ and $i+1$ of the links
and adds up a new link connecting points $i$ and $i+1$. If a loop is formed in the process, it must simply be
erased (loops are given a weight $1$ here, leading to the relation $e_i^2=e_i$).
Again, we use the integrable $R$ matrix, now corresponding to the choice
\eqn\integrab{ t(z,w)={q z-w\over q w-z} \qquad q=e^{2i\pi/3} }
Following again Sklyanin \SKLY,
we also introduce a boundary operator $K_i(z)$, whose action is diagonal at the points labelled $i=1$ or $N$, 
but whose effect is now to inverse the spectral parameter $z\to 1/z$ attached to that point, 
represented pictorially like in the Brauer case \pictobro\ (except that $-z$ is now replaced by $1/z$).
We still have the standard Yang-Baxter and unitarity relations (with multiplicative spectral parameters)
that read pictorially as in \ybunitbro, the boundary Yang-Baxter relation \bybro\ (with $-z$ and $-w$
replaced by $1/z$ and $1/w$ respectively),
and unitarity boundary relations at the leftmost and rightmost points:
\eqn\unitbountl{ K_1(z)K_1({1\over z})=I \qquad {\rm and}\qquad K_N(z)K_N({1\over z})=I }

The transfer matrix $T(t\vert z_1,z_2,\cdots ,z_N)$ of our model reads pictorially
exactly the same as in the Brauer case \pictotmbro, only the intersection between two oriented
lines carrying spectral parameters now correspond to the definition \rmat. It now acts on the vector 
space of (non-crossing) link patterns with $N$ points.
As a consequence of the Yang-Baxter and boundary Yang-Baxter equations, the transfer matrices at two 
distinct values of $t$ commute.

As before we denote by $\Psi^{(N)}(z_1,z_2,\cdots ,z_N)$ the common ground 
state vector of the $T$'s for fixed values of the $z_i$'s, satisfying \eigen.
As $T$ is a rational fraction of the $z_i$'s, we normalize $\Psi_N$ so that all its entries
are coprime polynomials of the $z_i$'s. We may view the entries of $\Psi^{(N)}$ as relative
probabilities of link pattern connections in random loop configurations with inhomogeneous probabilities
$(t_i,1-t_i)$ in the $i$-th column of the strip, and with $t_i=(qz_i-t)/(qt-z_i)$. This interpretation is 
again {\it stricto sensu} only valid
in the range of $z_i$'s leading to $t_i\in [0,1]$, in which case $\Psi^{(N)}$ is the Perron-Frobenius
eigenvector of $T$.

The remainder of this note is based on an empirical observation, which we conjecture to be
true, that for even $N=2n$, $\Psi^{(N)}$ defined above is a polynomial of total degree $3n(n-1)$ and partial
degree $2(n-1)$ in each variable. A similar property was proved in a rather indirect way in \DFZJ,
involving the details of the Bethe Ansatz solution of the corresponding integrable model. In the present case,
we believe such a proof should be within reach, although technically tedious, but we will content
ourselves with assuming the result. This property was the main difference between the
strategies of proof in the crossing and non-crossing periodic boundary loop models of 
Refs.\IDFZJ\ and \DFZJ,
the former appearing as more straightforward, as it does not require any
bound on the degree of the ground state vector.

An illustration is given
in Appendix B, where the entries of $\Psi^{(N)}$ are listed for the case $N=4$.

\subsec{Intertwining properties}

The intertwining relation \interl\ as well as its consequence \transpo\
still hold in the non-crossing case, with the appropriate 
definition \rmat\ of the $R$ matrix. 
When expressed in components, this translates into
\eqn\important{ \Delta_i \Psi_\pi^{(N)} =\sum_{\pi'\neq \pi\atop e_i \pi'=\pi} \Psi_{\pi'}^{(N)} }
where, for $i=1,2,...,N-1$, the operator $\Delta_i$ acts on functions 
$f\equiv f(z_1,z_2,\cdots ,z_N)$ as 
\eqn\divi{\Delta_i f={qz_i-z_{i+1}\over 1+q} \partial_i f}
for $i=1,2,...,N-1$, and with $\partial_i$ acting as in \actdf.
A first consequence of Eq.\transpo\ is that the entries $\Psi_\pi^{(N)}$ satisfy a suitably modified property
\item{\bf (P1)} If $z_j=q z_i$, and if the link pattern $\pi$ has no little arch connecting 
any pair of consecutive points between $i$ and $j$, then $\Psi_\pi^{(N)}$ vanishes.
\par
\noindent If $j=i+1$, this is easily deduced from the relation \transpo, by noting that 
$\Rc_{i,i+1}(z_i,qz_i)\propto e_i$. It is easily generalized to 
more distant points $i<j$ by considering suitable products of $R$ matrices
(see Ref.\DFZJ\ for a detailed proof in the periodic boundary case; the adaptation
to the open boundary case is straightforward).

\fig{A non-crossing link pattern (left) and its associated Dyck path (right). 
The box decomposition of the path is indicated, as well as the corresponding actions of $e_i$
on the fundamental link pattern made of consecutive arches connecting points
$2i-1$ and $2i$.}{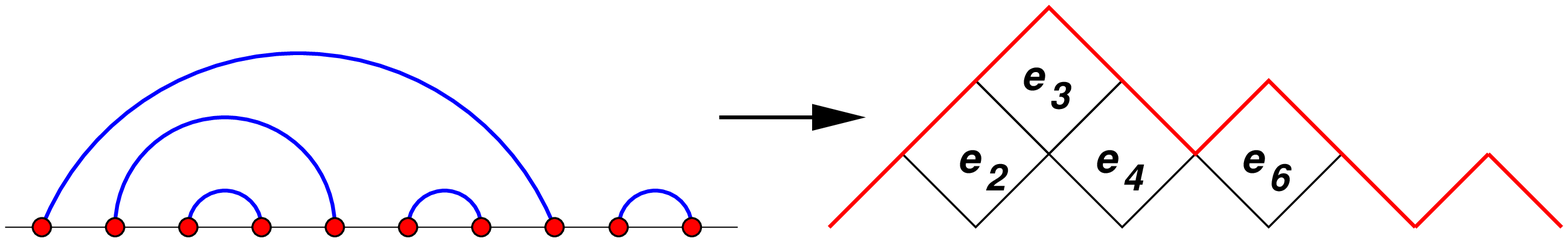}{12.cm}
\figlabel\dyck
For even $N=2n$,
these equations allow to determine all the entries of $\Psi^{(N)}$ in terms of that corresponding
to the link pattern $\pi_0$ with maximally nested arches, that connects points
$i$ and $2n+1-i$. Indeed, we may decompose any link pattern canonically into successive actions of $e_i$ 
on the "lowest" one, made of $n$ little arches connecting points $2i-1$ to $2i$. This is best seen in the
Dyck path formulation of link patterns, which are represented as paths on a square lattice as shown 
in Fig.\dyck. The path of a given link pattern is defined as follows. We visit the connected points
say from left to right, and parallelly draw a path with the rule that if we encounter a new arch
the path goes up one step, and if we encounter an arch already opened earlier, the path goes down one step.
The area below the path is then decomposed into square ``boxes", each of which corresponds
to an action with an operator $e_i$, whose index is the horizontal coordinate of the box, while the vertical
coordinate orders the successive actions. For instance, the decomposition
of  Fig.\dyck\ corresponds to acting with $e_3e_2e_4e_6$ on the fundamental link pattern made of 5 successive
little arches. With this formulation, it is easy to write 
down explicitly the antecedents $\pi'\neq \pi$ of a given link pattern $\pi$ under the action of $e_i$.
For these to exist, the Dyck path for $\pi$ must necessarily have a maximum at horizontal position $i$.
One obvious antecedent $\pi''$ is obtained by removing the box with this maximum. 
Others more subtle may arise from
adding a whole row of boxes. The important property here is that we may 
order antecedents by strict inclusion.
Indeed, the antecedent with the smallest number of boxes is $\pi''$, and is strictly contained in all others.
We may therefore express each new component in a triangular way with respect to strict inclusion. As an example
let us treat the case $N=6$, with $5$ link patterns, explicitly. We label $1,2,3,4,5$ the link patterns
and their corresponding Dyck paths
\eqn\lipanum{\figbox{12.cm}{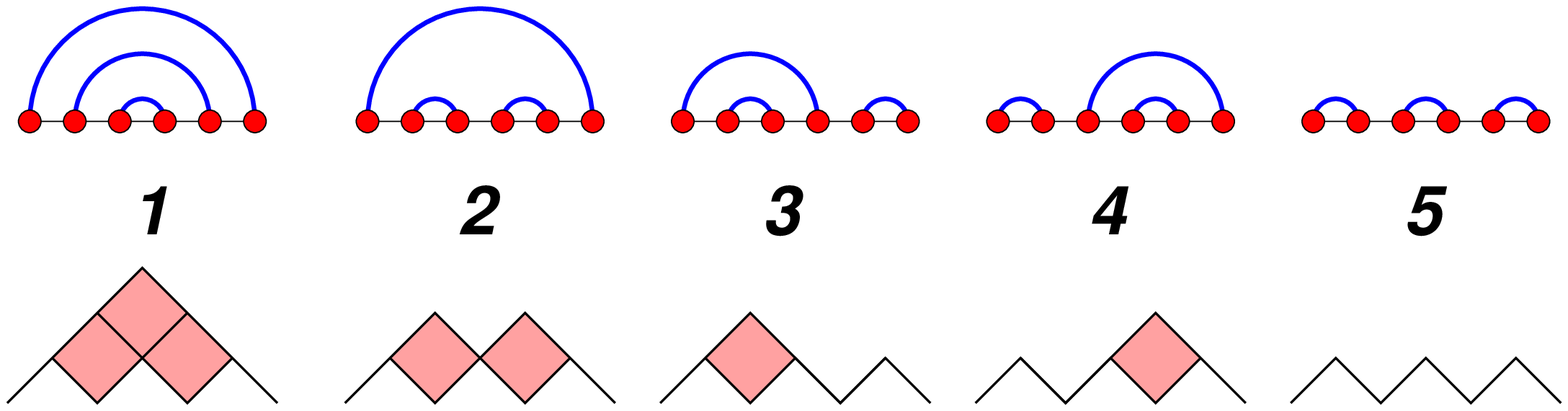} }
The relations \important\ allow us to express successively
\eqn\success{\eqalign{\Psi_2&=\Delta_3 \Psi_1\cr
\Psi_3&=\Delta_4 \Psi_2 -\Psi_1\cr
\Psi_4&=\Delta_2 \Psi_2 -\Psi_1\cr
\Psi_5&=\Delta_2 \Psi_3= \Delta_4 \Psi_4\cr}}
and we moreover have to write that 
\eqn\otrelTLA{\eqalign{
\Delta_5 \Psi_2&=\Psi_2\cr
\Delta_1 \Psi_4&=\Psi_2\cr
\Delta_1 \Psi_5&=\Psi_1+\Psi_3\cr
\Delta_3 \Psi_5&=\Psi_3+\Psi_4\cr
\Delta_5 \Psi_5&=\Psi_1+\Psi_4\cr}}
The compatibility between these equations implies a number of relations to be satisfied
by $\Psi_{\pi_0}^{(6)}\equiv \Psi_1$.
This construction also applies to the periodic case, upon cutting the link patterns between points $N$ and $1$
and opening them. Note finally that $\Delta_i$ are
degree preserving operators, hence the total and partial degrees of $\Psi_{\pi_0}^{(N)}$ 
are shared by all other entries of $\Psi^{(N)}$.

Let us finally mention the non-crossing loop model counterparts of Eqs.\intphibro\ and \interonebro.
We still denote by $\varphi_i$ the embedding of $LP_{2n-2}\to LP_{2n}$, that acts 
on a link pattern with $n-1$ arches
by inserting a little arch between points $i-1$ and $i$.
We have the following restriction/projection property:
if two neighboring parameters $z_i$ and $z_{i+1}$ are
such that $z_{i+1}=q\,z_i$, then
\eqn\intphi{T(t|z_1,\ldots,z_i,z_{i+1}=q\,z_i,\ldots,z_{2n})\, \varphi_i
=\varphi_i \,T(t|z_1,\ldots,z_{i-1},z_{i+2},z_{2n})}
Like in the Brauer case, this was proved in Ref.\DFZJ\ by explicitly commuting $\varphi_i$ through the product 
of two $R$ matrices at lines $i$ and $i+1$, and noting that $\Rc_{i,i+1}(z_i,qz_i)\propto e_i$.

Similarly, the equation \interonebro\ immediately translates into
\eqn\interone{T(t\vert z_1,z_2,\ldots ,z_N) K_1({1\over z_1}) 
=K_1({1\over z_1})T(t\vert {1\over z_1},z_2,\ldots ,z_N)}
with the same pictorial interpretation \interpic\ with $-z_1$ replaced by $1/z_1$. 

Finally, the reflection invariance of the system leads to the relation:
\eqn\inverv{(z_1z_2\ldots z_N)^{2(n-1)}\Psi^{(N)\, \rho}
({1\over z_N},{1\over z_{N-1}},\ldots ,{1\over z_1})=
\Psi^{(N)}(z_1,z_2,\ldots ,z_N)}
also obtained by implementing the condition that $\Psi^{(N)}$ 
has partial degree $2(n-1)$ in each variable $z_i$, and by using the reflection $\rho$ of non-crossing
link patterns.

\subsec{Solution for $\Psi^{(N)}_{\pi_0}$}

In the case of even $N=2n$,
applying the condition (i) to the maximally nested pattern $\pi_0$, with arches connecting points
$i$ and $2n-i$, we find that $\Psi_{\pi_0}^{(2n)}\equiv \Psi_{0}^{(2n)}$ must factor out the polynomial
$\prod_{1\leq i<j\leq n} (q z_i-z_j) \prod_{n+1\leq i<j\leq 2n}(q^2 z_j-z_i)$.
Moreover, the symmetry conditions \intertwo-\interfour\ imply more vanishing conditions, and henceforth
some extra polynomial factor $\prod_{2\leq j\leq n} (q^2z_1z_j-1)(qz_{n+j-1} z_N-1)$. This exhausts 
the partial degree $2(n-1)$ both in $z_1$ and $z_N$ of $\Psi_{0}^{(N)}$, hence we may write
for instance
\eqn\recpsi{\eqalign{ \Psi_{0}^{(N)}&(z_1,z_2,...,z_N)\cr
&=\prod_{j=2}^n(q z_1-z_j)(q^2z_1z_j-1) 
(q^2 z_N-z_{n+j-1})(qz_{n+j-1} z_N-1)\, 
A^{(N)}(z_2,...,z_{N-1})\cr}}
Now let us take $z_1\to 0$. The $R$-matrix elements involving $z_1$ in the transfer matrix $T$
reduce respectively to $-q^2 I\otimes I-q e_1$ and $-q I\otimes I-q^2 e_1$, whose product is $I\otimes I$. 
The net result
of taking $z_1\to 0$ is therefore to split $T$ into the identity acting at the point labelled $1$
and a similar transfer matrix acting on the points labelled $2,3,...,N$. Taking then $z_N\to 0$
now reduces the transfer matrix to that of size $N-2$, acting on the points labelled $2,3,...,N-1$.
As a result, $A^{(N)}$
is proportional to the maximally nested entry of the ground state vector at size $N-2$,
$\Psi^{(N-2)}_{0}(z_2,...,z_{N-1})$, and has the same partial and total degrees. 
We may now apply the condition (i) again, leading to more factors for $\Psi_{0}^{(N)}$.
Iterating this process, we exhaust all factors and finally reach the total degree $3n(n-1)$, with,
for even $N=2n$:
\eqn\psizero{ \Psi_{0}^{(2n)}(z_1,z_2,\cdots ,z_{2n})=\prod_{1\leq i<j\leq n} (q z_i-z_j)(q^2-z_iz_j)
\prod_{n+1\leq i<j\leq 2n}(q^2 z_j-z_i)(q-z_iz_j) }
As explained before, this determines in turn all components of $\Psi^{(N)}$ to be polynomials of the
same total and partial degrees.

As a by-product of the above discussion, the case of odd size $=2n-1$ is easily obtained from the 
case $N=2n$ by simply taking $z_{N}\to 0$. We then have that
\eqn\reczer{ \Psi^{(N-1)}(z_1,z_2,...,z_{N-1})=\lim_{z_N\to 0} 
\Psi^{(N)}(z_1,z_2,...,z_{N})}
Moreover, going from odd to even size gives
\eqn\reczertwo{ \Psi^{(N-2)}(z_1,z_2,...,z_{N-2})=\lim_{z_{N-1}\to 0} 
{\Psi^{(N-1)}(z_1,z_2,...,z_{N-1})\over z_1z_2...z_{N-2}} }

\subsec{Recursion relations}

As a consequence of the above intertwining properties, given a link pattern $\pi\in LP_{2n}$,
two situations may occur for a pair $(i,i+1)$ of consecutive points:

\noindent (i) the pattern $\pi$ has no arch joining $i$ to $i+1$,
in which case property (P1) implies that
\eqn\vanish{\Psi^{(N)}_\pi(z_1,\ldots,z_{N})\Big\vert_{z_{i+1}=q\,z_i}=0\ ;}

\noindent (ii) the pattern $\pi$ has a little arch joining $i$ to $i+1$, 
in which case
\eqn\recure{\eqalign{\Psi^{(N)}_\pi(z_1,\ldots,z_{N})\Big\vert_{z_{i+1}=q\,z_i}
&=\Psi^{(N-2)}_{\pi'}(z_1,\ldots,z_{i-1},z_{i+2},\ldots,z_{N})\cr
&\times \ \prod_{\scriptstyle k=1\atop\scriptstyle k\ne i,i+1}^{N} (q^2 z_i-z_k)(q^2z_iz_k-1)\cr}}
where $\pi'$ is the link pattern $\pi$ with the little arch $i$, $i+1$
removed ($\pi=\varphi_i\pi'$, $\pi'\in LP_{2n-2}$). This is readily obtained by applying 
Eq.\intphi\ to the vector $\Psi^{(N)}$ at $z_{i+1}=q\,z_i$. As a result, the
restricted $\Psi^{(N)}$ must be proportional to the projected $\Psi^{(N-2)}$, and the proportionality factor
is fixed by the value of $\Psi^{(N)}_{0}$ \psizero.

\subsec{Symmetries}

Like before, the intertwining properties of previous section lead straightforwardly to
symmetry properties for $\Psi^{(N)}$. Indeed, applying the relation \intphi\ on the
vector $\Psi^{(N)}({1\over z_1},z_2,\ldots ,z_N)$, we find that the latter must be proportional to
$\Psi^{(N)}(z_1,z_2,\ldots ,z_N)$, with the result:
\eqn\intertwo{z_1^{4(n-1)}\Psi^{(N)}({1\over z_1},z_2,\ldots ,z_N)=\Psi^{(N)}(z_1,z_2,\cdots ,z_N)}
The same reasoning at the other end with the space labelled $N$ leads to the condition
\eqn\interfour{z_N^{4(n-1)}\Psi^{(N)}(z_1,z_2,\ldots z_{N-1},{1\over z_N})=\Psi^{(N)}(z_1,z_2,\ldots ,z_N)}
In both equations, we have used the assumed fact that $\Psi^{(N)}$ has partial degree $2(n-1)$
in each variable $z_i$.

As before, let us consider the sum over all components of $\Psi^{(N)}$, namely
\eqn\defZ{ Z^{(N)}(z_1,...,z_N)=\sum_{\pi\in LP_N} \Psi_\pi^{(N)}(z_1,z_2,...,z_N)=v_N\cdot \Psi^{(N)}}
where $v_N$ is the vector with all entries equal to $1$. This vector also satisfies
\eqn\vecsat{ v_N e_i=v_N\qquad {\rm and}\qquad  v_N{\check R}_{i,i+1}(z_i,z_{i+1}) =v_N }
hence applying $v_N$ to Eq.\transpo, we immediately get that 
$\tau_i Z^{(N)} =Z^{(N)}$, $i=1,2,...,N-1$, henceforth $Z$ is symmetric in the $z_i$'s.
Moreover, applying $v_N$ to Eq.\intertwo, we get
\eqn\otsym{ z_1^{4(n-1)} Z^{(N)}({1\over z_1},z_2,...,z_N)=Z^{(N)}(z_1,z_2,...,z_N)}
hence $Z^{(N)}$ is symmetric and reciprocal in each of the $z$'s.

\subsec{Sum rule}

For even $N=2n$, we have
\eqn\sumrul{\eqalign{Z^{(N)}(z_1,..,z_{2n})&={\prod_{i,j=1}^n 
(z_i^2+z_iz_{j+n}+z_{j+n}^2)(1+z_iz_{j+n}+z_i^2z_{j+n}^2)
\over \prod_{1\leq i<j \leq n} (z_i-z_j)(1-z_iz_j)(z_{i+n}-z_{j+n})(1-z_{i+n}z_{j+n})}\cr
&\times
\det\left({1\over z_i^2+z_iz_{j+n}+z_{j+n}^2}
{1\over 1+z_iz_{j+n}+z_i^2z_{j+n}^2}\right)_{1\leq i,j\leq n}\cr}}
Remarkably, this coincides with the partition function $Z_{UASM}(z_1,..,z_n;z_{n+1},...,z_{2n})$
introduced in Ref.\KUP.
The proof of Eq.\sumrul\ parallels exactly that of Eq.\sumrulbrau, 
proceeding by induction on $n$, and makes use of the recursion relations Eq.\recure, 
as well as of the symmetries of $Z^{(N)}$.

We also have
\eqn\othsumrul{\eqalign{Z^{(N)}(z_1,..,z_{2n})^2&=\prod_{1\leq i<j \leq 2n}{z_i^2+z_iz_j+z_j^2\over z_i-z_j}
{1+z_iz_j+z_i^2z_j^2\over 1-z_iz_j} \cr
&\times {\rm Pf}
\left({z_i-z_j\over z_i^2+z_iz_j+z_j^2}{1-z_iz_j\over 1+z_iz_j+z_i^2z_j^2}\right)_{1\leq i,j\leq N}\cr}}
also proved by induction on $n$. The latter expression has the advantage of
being explicitly symmetric in the $z$'s.

In the homogeneous limit where all $z_i$'s tend to 1, we find that for $N=2n$, 
$Z^{(2n)}(1,1...,1)$ is $3^{n(n-1)}$ times the total number of U-symmetric ASM's of size $(2n)\times (2n)$
discussed in Ref.\KUP, itself identical to that of vertically symmetric ASM's of size $(2n+1)\times (2n+1)$.

\newsec{Conclusion}

In this paper, we have derived sum rules for the ground state vector
of the inhomogeneous crossing and non-crossing O(1)
loop models on a semi-infinite strip. As opposed to the crossing case where the result
is rigorous and proved completely, we have made in the non-crossing case
a reasonable but crucial assumption on the degree
of the ground state vector as a polynomial of the inhomogeneities $z_i$. The completion of the latter
proof would presumably involve invoking the algebraic Bethe Ansatz solution of the XXZ spin chain
with open boundaries, in much the same spirit as in Ref.\DFZJ. We have rather chosen here to concentrate 
on the various properties of this ground state vector, for which we gave an explicit step-by-step 
construction by acting on a fundamental component with local divided difference operators, in
order to generate all other entries of the vector.
In this respect, it might be possible to unify both crossing and non-crossing cases by deriving a proof
uniquely based on the main relations induced by Eq.\transpo, and that only involve the interplay
between the symmetric group action on spectral parameters and the $R$-matrix of the integrable system.
In order to do this, and by analogy with the crossing case,
one should be able to check that the solution of the non-crossing case \psizero\ actually satisfies
all constraints inherited from the compatibility of all equations \important, as well as the
boundary reflection properties \intertwo-\interfour. By a uniqueness argument, this would by-pass
our approach, which assumes the value of the degree of $\Psi^{(N)}$. This is a problem for future 
work. 
Note finally that such a construction, both in the crossing and non-crossing cases, 
should be instrumental in 
trying to prove variants of the so-called Razumov-Stroganov conjectures \RS.

Our approach allows in particular to compute the entries of the ground state vector in the homogeneous limit,
where it may be identified with the ground state vector of the Hamiltonian of a suitable quantum chain,
expressed as a particular weighted sum of generators of the Brauer (resp. Temperley-Lieb) algebra
for the crossing (resp. non-crossing) case, acting on crossing (resp. non-crossing) link patterns
(see Refs.\BRAU\ and \RS\ for explicit expressions). 
As an outcome of our calculation, we show that these 
entries may be picked to be non-negative integers, summing to specific numbers as given by 
\homoZbro-\oddhomobro\ for strips of even/odd size in the crossing case, and to the total number 
of vertically symmetric alternating sign matrices $A_V(2n+1)$ of \KUP\ for strips of even size $N=2n$.
It is important to note that, as opposed to the standard case
where the entries of the homogeneous ground state vector are normalized so that the smallest one 
is $1$ after division by their GCD (c.f. \IDFZJ\ \DFZJ), the smallest entries in the open crossing 
case are not $1$, but form themselves a quite intriguing sequence \valpsizerobro, as derived from
the homogeneous limit of the relation \cloP.

While the numbers $A_V(2n+1)$ have been given extensive combinatorial interpretations, that of the numbers
\homoZbro-\oddhomobro\ is still elusive. 
Such an interpretation was suggested in Ref.\BRAU\ for the cylinder case, by noticing 
and conjecturing that some entries of
the homogeneous ground state vector of the crossing loop model with periodic boundaries
actually matched degrees of varieties related to the commuting variety \Kn.
This was further proved in \IDFZJ\ and extended in \KZJ, where all the components of the ground state
vector were interpreted as the multidegrees of the components of a matrix variety.  It is natural to hope that
the numbers \homoZbro-\oddhomobro\ actually count the total degrees of some matrix varieties,
still to be found.
In this respect, the partial sum rule $W^{(N)}$ in the permutation sector \sumrulperbro\ leading to the numbers 
\homoW\ seems to indicate, like in the periodic case, that the corresponding components of the (yet unknown)
matrix variety form a complete intersection, whose multidegree has the factorized form \sumrulperbro.
We could also hope that the total multidegree, as given by Eq.\sumrulbrau, may be alternatively obtained
like in \KZJ\ as the result of a ``volume" matrix integral over the putative matrix variety.

\bigskip
\leftline{\bf Acknowledgments}
We acknowledge many interesting and stimulating discussions with P. Zinn-Justin and J.-B. Zuber.
This research was partly supported by the european network ENIGMA, grant MRTN-CT-2004-5652
and of the GEOCOMP project (ACI ``Masse de donn\'ees").

\vfill\eject

\appendix{A}{Entries of the ground state vector in the dense O(1) crossing loop model 
with open boundaries}

In this appendix, we give the entries of $\Psi^{(N)}$ and their sum $Z^{(N)}(z_1,z_2,z_3,z_4)$
in the Brauer loop case for $N=4$, as well as the entry $\Psi^{(N)}_{\pi_0}$ for $N=6$.
In the notations of Sect.2.3, and
for $N=4$, the three components of $\Psi^{(4)}$ read:
\eqn\psifourbrau{\eqalign{ \Psi^{(4)} \raise-.4cm\hbox{\epsfxsize=1cm\epsfbox{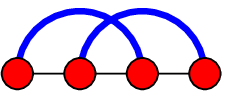}} &=
a_{1,2}b_{1,2}a_{2,3}b_{3,4}c_{3,4}
(5+3z_2-3z_3-2z_2z_3-z_1^2-z_4^2+(z_1^2-z_4^2)(z_2+z_3))\cr
\Psi^{(4)} \raise-.4cm\hbox{\epsfxsize=1cm\epsfbox{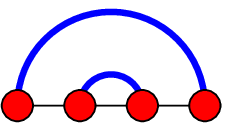}} &=
a_{1,2}b_{1,2}a_{3,4}c_{3,4}
(11-3z_1^2+8z_2+z_2^2-8z_3+2z_1^2z_3-8z_2z_3-2 z_2^2 z_3 \cr
&+z_3^2+2z_2z_3^2-3z_4^2-2z_2z_4^2-(z_1^2-z_3^2)(z_2^2-z_4^2))\cr
\Psi^{(4)} \raise-.4cm\hbox{\epsfxsize=1cm\epsfbox{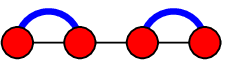}} &=
a_{2,3}\big(23 - 10 z_1^2 + 3 z_1^4 - 7 z_2 + 18 z_1^2 z_2 - 
3 z_1^4 z_2 - 11 z_2^2 + 4 z_1^2 z_2^2 - z_1^4 z_2^2 +3 z_2^3\cr
&-4 z_1^2 z_2^3+7z_3+10 z_1^2z_3-z_1^4z_3+2z_2 z_3+ 
16 z_1^2 z_2 z_3-2z_1^4z_2z_3-z_2^2z_3 -4 z_1^2 z_2^2 z_3 \cr
&-4z_2^3z_3-11z_3^2+9z_1^2 z_3^2-2z_1^4z_3^2+z_2z_3^2+ 
z_1^2 z_2 z_3^2 + 3 z_2^2 z_3^2+z_1^2 z_2^2 z_3^2 - z_2^3 z_3^2- 3 z_3^3 \cr
&- z_1^2 z_3^3 - 
4 z_2 z_3^3 - 2 z_1^2 z_2 z_3^3 + z_2^2 z_3^3 
+ 2 z_2^3 z_3^3 -10 z_4^2-11 z_1^2 z_4^2+ z_1^4 z_4^2 - 10 z_2 z_4^2 \cr
&+z_1^2 z_2 z_4^2 + 9 z_2^2 z_4^2 + 
z_1^2 z_2^2 z_4^2 + z_2^3 z_4^2- 18 z_3 z_4^2- z_1^2 z_3 z_4^2+ 16 z_2 z_3 z_4^2 + 
6 z_1^2 z_2 z_3 z_4^2 \cr
&- z_2^2 z_3 z_4^2- 2 z_2^3 z_3 z_4^2+4 z_3^2 z_4^2+ z_1^2 z_3^2 z_4^2 +4 z_2 z_3^2 z_4^2
+z_2^2 z_3^2 z_4^2+4 z_3^3 z_4^2+ 3 z_4^4 \cr
&+ z_1^2 z_4^4 + 
z_2 z_4^4-2z_2^2 z_4^4+3 z_3z_4^4-2z_2z_3z_4^4-z_3^2 z_4^4\cr
&+(z_2+z_3)(z_1^2-z_3^2)(z_1^2-z_4^2)(z_2^2-z_4^2)\big)\cr}}
These may be obtained by explicitly solving the eigenvector equation \eigen. Alternatively,
we have computed $\Psi_{0}^{(4)}$ in the text using \inducpsi-\instfour, with the result
$\Psi_{0}^{(4)}=a_{1,2}b_{1,2}a_{2,3}b_{3,4}c_{3,4}P_0^{(4)}$. The other components
read simply
\eqn\simother{\eqalign{\Psi^{(4)} \raise-.4cm\hbox{\epsfxsize=1cm\epsfbox{lp2-4.eps}} &=
\Theta_1 \ \Psi^{(4)} \raise-.4cm\hbox{\epsfxsize=1cm\epsfbox{lp1-4.eps}}\cr
\Psi^{(4)} \raise-.4cm\hbox{\epsfxsize=1cm\epsfbox{lp3-4.eps}} &=
\Theta_2 \ \Psi^{(4)} \raise-.4cm\hbox{\epsfxsize=1cm\epsfbox{lp1-4.eps}}
\cr}}

The components \psifourbrau\ sum to:
\eqn\psisumfourbrau{\eqalign{Z^{(4)}(z_1,z_2,z_3,z_4)&=
39-30z_1^2+7z_1^4-30z_2^2+14z_1^2z_2^2-4z_1^4z_2^2+7z_2^4-4z_1^2z_2^4\cr
&+ z_1^4z_2^4-30z_3^2+14z_1^2z_3^2-4z_1^4z_3^2+14z_2^2z_3^2+12z_1^2 z_2^2z_3^2-z_1^4z_2^2z_3^2 
\cr
&-4z_2^4z_3^2-z_1^2z_2^4z_3^2+7z_3^4-4z_1^2z_3^4+z_1^4z_3^4-4 z_2^2z_3^4-z_1^2 z_2^2z_3^4+z_2^4z_3^4-30z_4^2
\cr
&+14z_1^2z_4^2-4z_1^4z_4^2+14 z_2^2z_4^2+12z_1^2z_2^2z_4^2-z_1^4z_2^2z_4^2-4z_2^4z_4^2-z_1^2z_2^4z_4^2+14z_3^2z_4^2
\cr
&+12z_1^2z_3^2z_4^2-z_1^4z_3^2z_4^2+12z_2^2z_3^2z_4^2+6z_1^2z_2^2z_3^2z_4^2-z_2^4z_3^2z_4^2-4z_3^4z_4^2 \cr
&-z_1^2z_3^4z_4^2-z_2^2z_3^4z_4^2+7z_4^4-4z_1^2z_4^4+z_1^4z_4^4-4 z_2^2z_4^4-z_1^2 z_2^2z_4^4+ z_2^4z_4^4\cr
&-4z_3^2z_4^4-z_1^2z_3^2z_4^4-z_2^2z_3^2z_4^4+z_3^4z_4^4\cr}}
It takes only a few seconds for any formal manipulation software to check that this quantity
indeed coincides with the Pfaffian expression of Eq.\sumrulbrau.

We also display the value of the component $\Psi^{(6)}_{0}$ for $N=6$, as obtained from the 
formula \cloP\ and with extensive use of the modified
Leibniz formula \modilei:
\eqn\psisixbrau{\eqalign{ &\Psi^{(6)} \raise-.4cm\hbox{\epsfxsize=1.5cm\epsfbox{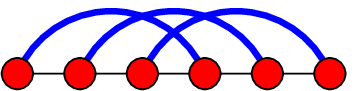}}=
a_{1,2}b_{1,2}a_{1,3}b_{1,3}a_{2,3}b_{2,3}a_{2,4}a_{3,4}a_{3,5}a_{4,5}c_{4,5}a_{4,6}c_{4,6}
a_{5,6}c_{5,6}\times\cr
&\times
\Big(a_{1, 6}b_{1, 6}a_{2, 6}b_{2, 6}b_{3, 4}b_{3, 5}(a_{1, 5} b_{1, 5} b_{2, 4} + 2 c_{2, 5}(1 + b_{5, 4}))\cr
&+4a_{1, 6}a_{2, 6}b_{3, 4}c_{3, 6}(a_{1, 5} b_{2, 4} c_{2, 6}+2 b_{1, 6} (1 + b_{5, 4}))\cr
&+4c_{3, 5}c_{3, 6}(1 + b_{4, 6})(1 + b_{5, 6})(a_{1, 5} b_{1, 6} b_{2, 4} + 2 c_{2, 6}(1 + b_{5, 4}))\cr
&+2c_{3, 5}c_{3, 6}(1 + b_{4, 6})(a_{1, 5} a_{2, 6} b_{1, 5}b_{2, 4} c_{2, 6}
+2 a_{1, 5} b_{1, 5} b_{1, 6} (a_{1, 6}- b_{2, 4}) + 
2 a_{1, 6} b_{1, 4} b_{1, 6} (b_{2, 5}-a_{1, 5}))\cr
&+2c_{3, 6}b_{3, 4}(1 + b_{5, 6})(a_{1, 5} a_{1, 6} b_{1, 5} b_{1, 6} b_{2, 4} + 
2 b_{1, 4}a_{2, 6} c_{2, 6}(b_{2, 5}-a_{1, 5}) + 
2 a_{1, 5}b_{1, 5} c_{2, 6} (1 + b_{4, 6}))
\Big)\cr}}
As all $a,b,c$'s tend to $1$ in the homogeneous limit where all $z\to 0$, we read off \psisixbrau\
that $\Psi_{0}^{(6)}(0,0,0,0,0,0)=129$. As expected for the case of arbitrary even $N$,
the result for $\Psi_{0}^{(6)}$ is an integer 
linear combination of products of $a,b,c$'s, with coefficients $\pm$ powers of $2$.

\vfill\eject

\appendix{B}{Entries of the ground state vector in the dense O(1) loop model with open boundaries}

In this appendix, we give the entries of $\Psi^{(N)}$ and their sum $Z^{(N)}$
in the non-crossing loop case for $N=4$.
In the notations of Sect.3, and
for $N=4$, the two components of $\Psi^{(4)}$ read:
\eqn\psifourbrau{\eqalign{ \Psi^{(4)} \raise-.4cm\hbox{\epsfxsize=1.5cm\epsfbox{lp2-4.eps}} &=
(qz_1-z_2)(q^2z_1z_2-1)(q^2z_3-z_4)(qz_3z_4-1)\cr
\Psi^{(4)} \raise-.4cm\hbox{\epsfxsize=1.5cm\epsfbox{lp3-4.eps}} &=
(qz_2-z_3)\big((q^2z_1-z_4)(1+z_1z_2z_3z_4)+(q^2z_4-z_1)(z_1z_4+z_2z_3)\cr
&+(q^2-1)z_1z_4(z_2+z_3)\big)\cr}}
These components sum to
\eqn\sumpsitl{\eqalign{ Z^{(4)}(z_1,z_2,z_3,z_4)&= (1+z_1z_2z_3z_4)
(z_1z_2+z_1z_3+z_2z_3+z_1z_4+z_2z_4+z_3z_4)\cr
&+(z_1 +z_2)(z_3+z_4)(z_1z_2+z_3z_4)+(z_1z_3+z_2z_4)(z_1z_4+z_2z_3)\cr
&+3 z_1z_2z_3z_4\cr}}
which may easily be checked against Eqs.\sumrul\ and \othsumrul.

\listrefs
\end

The polynomial $P^{(6)}$, of degree $9$, reads
$$\eqalign{
P&= 129-54 z_1^2 + 5 z_1^4 + 79\ z_2 + 6 z_1^2 z_2 - 
        5 z_1^4 z_2 - 29 z_2^2 + 14 z_1^2 z_2^2 - 
        z_1^4 z_2^2 - 19 z_2^3 + 2 z_1^2 z_2^3 \cr
	&+ 
        z_1^4 z_2^3 + 158 z_3 - 28 z_1^2 z_3 - 2 z_1^4 z_3 + 
        98 z_2 z_3 + 28 z_1^2 z_2 z_3 + 2 z_1^4 z_2 z_3 + 
        2 z_2^2 z_3 - 4 z_1^2 z_2^2 z_3 + 
        2 z_1^4 z_2^2 z_3 - 2 z_2^3 z_3 \cr
	&+ 
        4 z_1^2 z_2^3 z_3 - 2 z_1^4 z_2^3 z_3 + 49 z_3^2 + 
        2 z_1^2 z_3^2 - 3 z_1^4 z_3^2 + 31 z_2 z_3^2 + 
        14 z_1^2 z_2 z_3^2 + 3 z_1^4 z_2 z_3^2 + 
        11 z_2^2 z_3^2 + 6 z_1^2 z_2^2 z_3^2 \cr
	&- 
        z_1^4 z_2^2 z_3^2 + 5 z_2^3 z_3^2 + 
        10 z_1^2 z_2^3 z_3^2 + z_1^4 z_2^3 z_3^2 - 
        158 z_4 + 80 z_1^2 z_4 - 10 z_1^4 z_4 - 103 z_2 z_4 + 
        10 z_1^2 z_2 z_4 + 5 z_1^4 z_2 z_4 \cr
	&+ 
        38 z_2^2 z_4 - 16 z_1^2 z_2^2 z_4 + 
        2 z_1^4 z_2^2 z_4 + 27 z_2^3 z_4 - 
        2 z_1^2 z_2^3 z_4 - z_1^4 z_2^3 z_4 - 
        201 z_3 z_4 + 58 z_1^2 z_3 z_4 - z_1^4 z_3 z_4 \cr
	&- 
        133 z_2 z_3 z_4 - 14 z_1^2 z_2 z_3 z_4 + 
        3 z_1^4 z_2 z_3 z_4 + 5 z_2^2 z_3 z_4 + 
        14 z_1^2 z_2^2 z_3 z_4 - 3 z_1^4 z_2^2 z_3 z_4 + 
        9 z_2^3 z_3 z_4 + 6 z_1^2 z_2^3 z_3 z_4 \cr
	&+ 
        z_1^4 z_2^3 z_3 z_4 - 65 z_3^2 z_4 + 
        6 z_1^2 z_3^2 z_4 + 3 z_1^4 z_3^2\ z_4 - 
        44 z_2 z_3^2 z_4 - 12 z_1^2 z_2 z_3^2\ z_4 - 
        11 z_2^2 z_3^2 z_4 + 2 z_1^2 z_2^2 z_3^2 z_4 \cr
	&+ 
        z_1^4 z_2^2 z_3^2 z_4 - 4 z_2^3 z_3^2 z_4 - 
        4 z_1^2 z_2^3 z_3^2 z_4 + 49 z_4^2 - 
        30 z_1^2 z_4^2 + 5 z_1^4 z_4^2 + 34 z_2 z_4^2 - 
        10 z_1^2 z_2 z_4^2 - 13 z_2^2 z_4^2 \cr
	&+ 
        6 z_1^2 z_2^2 z_4^2 - z_1^4 z_2^2 z_4^2 - 
        10 z_2^3 z_4^2 + 2 z_1^2 z_2^3 z_4^2 + 
        65 z_3 z_4^2 - 28 z_1^2 z_3 z_4^2 + 
        3 z_1^4 z_3 z_4^2 + 46 z_2 z_3 z_4^2 - 
        6 z_1^2 z_2 z_3 z_4^2 - 5 z_2^2 z_3 z_4^2 - 
        4 z_1^2 z_2^2 z_3 z_4^2 \cr
	&+ 
        z_1^4 z_2^2 z_3 z_4^2 - 6 z_2^3 z_3 z_4^2 - 
        2 z_1^2 z_2^3 z_3 z_4^2 + 22 z_3^2 z_4^2 - 
        6 z_1^2 z_3^2 z_4^2 + 16 z_2 z_3^2 z_4^2 + 
        2 z_2^2 z_3^2 z_4^2 - 
        2 z_1^2 z_2^2 z_3^2 z_4^2 - 79 z_5 + 
        34 z_1^2 z_5 - 3 z_1^4 z_5 - 53 z_2 z_5 \cr
	&+ 
        2 z_1^2 z_2 z_5 + 3 z_1^4 z_2 z_5 + 21 z_2^2 z_5 - 
        6 z_1^2 z_2^2 z_5 + z_1^4 z_2^2z_5 + 
        15 z_2^3 z_5 + 2 z_1^2 z_2^3 z_5 - 
        z_1^4 z_2^3 z_5 - 103 z_3 z_5 + 22 z_1^2 z_3 z_5 + 
        z_1^4 z_3 z_5 - 73 z_2 z_3 z_5 - 
        6 z_1^2 z_2 z_3 z_5 - z_1^4 z_2 z_3 z_5 \cr
	&+ 
        7 z_2^2 z_3 z_5 + 10 z_1^2 z_2^2 z_3 z_5 - 
        z_1^4 z_2^2 z_3 z_5 + 9 z_2^3 z_3 z_5 + 
        6 z_1^2 z_2^3 z_3 z_5 + z_1^4 z_2^3 z_3 z_5 - 
        34 z_3^2 z_5 + 2 z_1^4 z_3^2 z_5 - 
        26 z_2 z_3^2 z_5 - 4 z_1^2 z_2 z_3^2 z_5 - 
        2 z_1^4 z_2 z_3^2 z_5 - 4 z_2^2 z_3^2 z_5 \cr
	&+ 
        4 z_1^2 z_2^2 z_3^2 z_5 + 98 z_4 z_5 - 
        48 z_1^2 z_4 z_5 + 6 z_1^4 z_4 z_5 + 
        73 z_2 z_4 z_5 - 6 z_1^2 z_2 z_4 z_5 - 
        3 z_1^4 z_2 z_4 z_5 - 22 z_2^2 z_4 z_5 + 
        16 z_1^2 z_2^2 z_4 z_5 - 2 z_1^4 z_2^2 z_4 z_5 - 
        19 z_2^3 z_4 z_5 + 2 z_1^2 z_2^3 z_4 z_5 \cr
	&+ 
        z_1^4 z_2^3 z_4 z_5 + 133 z_3 z_4 z_5 - 
        38 z_1^2 z_3 z_4 z_5 + z_1^4 z_3 z_4 z_5 + 
        106 z_2 z_3 z_4 z_5 + 8 z_1^2 z_2 z_3 z_4 z_5 - 
        2 z_1^4 z_2 z_3 z_4 z_5 - 3 z_2^2 z_3 z_4 z_5 + 
        2 z_1^2 z_2^2 z_3 z_4 z_5 + 
        z_1^4 z_2^2 z_3 z_4 z_5 - 12 z_2^3 z_3 z_4 z_5 \cr
	&- 
        4 z_1^2 z_2^3 z_3 z_4 z_5 + 46 z_3^2 z_4 z_5 - 
        4 z_1^2 z_3^2 z_4 z_5 - 2 z_1^4 z_3^2 z_4 z_5 + 
        40 z_2 z_3^2 z_4 z_5 + 
        8 z_1^2 z_2 z_3^2 z_4 z_5 + 
        8 z_2^2 z_3^2 z_4 z_5 - 31 z_4^2 z_5 + 
        18 z_1^2 z_4^2 z_5 - 3 z_1^4 z_4^2 z_5 \cr
	&- 
        26 z_2 z_4^2 z_5 + 6 z_1^2 z_2 z_4^2 z_5 + 
        5 z_2^2 z_4^2 z_5 - 6 z_1^2 z_2^2 z_4^2 z_5 + 
        z_1^4 z_2^2 z_4^2 z_5 + 6 z_2^3 z_4^2 z_5 - 
        2 z_1^2 z_2^3 z_4^2 z_5 - 44 z_3 z_4^2 z_5 + 
        18 z_1^2 z_3 z_4^2 z_5 - 2 z_1^4 z_3 z_4^2 z_5 - 
        40 z_2 z_3 z_4^2 z_5 \cr
	&+ 
        4 z_1^2 z_2 z_3 z_4^2 z_5 - 
        2 z_2^2 z_3 z_4^2z_5 - 
        2 z_1^2 z_2^2 z_3 z_4^2 z_5 + 
        4 z_2^3 z_3 z_4^2 z_5 - 16 z_3^2 z_4^2 z_5 + 
        4 z_1^2 z_3^2 z_4^2 z_5 - 
        16 z_2 z_3^2 z_4^2 z_5 \cr
	&- 
        4 z_2^2 z_3^2 z_4^2 z_5 - 29 z_5^2 + 
        9 z_1^2 z_5^2 - 21 z_2 z_5^2 + z_1^2 z_2 z_5^2 + 
        5 z_2^2 z_5^2 - z_1^2 z_2^2 z_5^2 + 
        5 z_2^3 z_5^2 - z_1^2 z_2^3 z_5^2 - 38 z_3 z_5^2 + 
        6 z_1^2 z_3 z_5^2 - 22 z_2 z_3 z_5^2 - 
        10 z_1^2 z_2 z_3 z_5^2 - 2 z_2^2 z_3 z_5^2 \cr
	&+ 
        2 z_1^2 z_2^2 z_3 z_5^2 - 2 z_2^3 z_3 z_5^2 + 
        2 z_1^2 z_2^3 z_3 z_5^2 - 13 z_3^2 z_5^2 + 
        z_1^2 z_3^2 z_5^2 - 5 z_2 z_3^2 z_5^2 - 
        7 z_1^2 z_2 z_3^2 z_5^2 - 3 z_2^2 z_3^2 z_5^2 - 
        z_1^2 z_2^2 z_3^2 z_5^2 - 3 z_2^3 z_3^2 z_5^2 - 
        z_1^2 z_2^3 z_3^2 z_5^2 - 2 z_4 z_5^2 \cr
	&+ 
        7 z_2 z_4 z_5^2 - 5 z_1^2 z_2 z_4 z_5^2 + 
        2 z_2^2 z_4 z_5^2 - 3 z_2^3 z_4 z_5^2 + 
        z_1^2 z_2^3 z_4 z_5^2 + 5 z_3 z_4 z_5^2 - 
        5 z_1^2 z_3 z_4 z_5^2 + 3 z_2 z_3 z_4 z_5^2 - 
        3 z_1^2 z_2 z_3 z_4 z_5^2 - 
        z_2^2 z_3 z_4 z_5^2 \cr
	&+ 
        z_1^2 z_2^2 z_3 z_4 z_5^2 + 
        z_2^3 z_3 z_4 z_5^2 - 
        z_1^2 z_2^3 z_3 z_4 z_5^2 + 5 z_3^2 z_4 z_5^2 - 
        3 z_1^2 z_3^2 z_4 z_5^2 - 
        2 z_2 z_3^2 z_4 z_5^2 - z_2^2 z_3^2 z_4 z_5^2 - 
        z_1^2 z_2^2 z_3^2 z_4 z_5^2 \cr
	&+ 
        2 z_2^3 z_3^2 z_4 z_5^2 + 11 z_4^2 z_5^2 - 
        5 z_1^2 z_4^2 z_5^2 + 4 z_2 z_4^2 z_5^2 - 
        3 z_2^2 z_4^2 z_5^2 + z_1^2 z_2^2 z_4^2 z_5^2 + 
        11 z_3 z_4^2 z_5^2 - 3 z_1^2 z_3 z_4^2 z_5^2 + 
        8 z_2 z_3 z_4^2 z_5^2 + z_2^2 z_3 z_4^2 z_5^2 - 
        z_1^2 z_2^2 z_3 z_4^2 z_5^2 \cr
	&+ 
        2 z_3^2 z_4^2 z_5^2 + 4 z_2 z_3^2 z_4^2 z_5^2 + 
        2 z_2^2 z_3^2 z_4^2 z_5^2 + 19 z_5^3 - 
        7 z_1^2 z_5^3 + 15 z_2 z_5^3 - 3 z_1^2 z_2 z_5^3 - 
        5 z_2^2 z_5^3 + z_1^2 z_2^2 z_5^3 - 
        5 z_2^3 z_5^3 + z_1^2 z_2^3 z_5^3 + 27 z_3 z_5^3 - 
        7 z_1^2 z_3 z_5^3 + 19 z_2 z_3 z_5^3 \cr
	&+ 
        z_1^2 z_2 z_3 z_5^3 - 3 z_2^2 z_3 z_5^3 - 
        z_1^2 z_2^2 z_3 z_5^3 - 3 z_2^3 z_3 z_5^3 - 
        z_1^2 z_2^3 z_3 z_5^3 + 10 z_3^2 z_5^3 - 
        2 z_1^2 z_3^2 z_5^3 + 6 z_2 z_3^2 z_5^3 + 
        2 z_1^2 z_2 z_3^2 z_5^3 - 2 z_4 z_5^3 - 
        9 z_2 z_4 z_5^3 + 3 z_1^2 z_2 z_4 z_5^3 \cr
	&- 
        2 z_2^2 z_4 z_5^3 + 3 z_2^3 z_4 z_5^3 - 
        z_1^2 z_2^3 z_4 z_5^3 - 9 z_3 z_4 z_5^3 + 
        3 z_1^2 z_3 z_4 z_5^3 - 12 z_2 z_3 z_4 z_5^3 + 
        2 z_1^2 z_2 z_3 z_4 z_5^3 - 
        z_2^2 z_3 z_4 z_5^3 - 
        z_1^2 z_2^2 z_3 z_4 z_5^3 \cr
	&+ 
        2 z_2^3 z_3 z_4 z_5^3 - 6 z_3^2 z_4 z_5^3 + 
        2 z_1^2 z_3^2 z_4 z_5^3 - 
        4 z_2 z_3^2 z_4 z_5^3 - 5 z_4^2 z_5^3 + 
        3 z_1^2 z_4^2 z_5^3 + 3 z_2^2 z_4^2 z_5^3 - 
        z_1^2 z_2^2 z_4^2 z_5^3 - 4 z_3 z_4^2 z_5^3 + 
        2 z_1^2 z_3 z_4^2 z_5^3 + 
        2 z_2^2 z_3 z_4^2 z_5^3 - 54 z_6^2 \cr}$$
	
\eqn\polbrau{\eqalign{&+ 
        19 z_1^2 z_6^2 - z_1^4 z_6^2 - 34 z_2 z_6^2 - 
        3 z_1^2 z_2 z_6^2 + z_1^4 z_2 z_6^2 + 
        9 z_2^2 z_6^2 - 5 z_1^2 z_2^2 z_6^2 + 
        7 z_2^3 z_6^2 - 3 z_1^2 z_2^3 z_6^2 - 
        80 z_3 z_6^2 + 16 z_1^2 z_3 z_6^2 - 
        48 z_2 z_3 z_6^2 - 16 z_1^2 z_2 z_3 z_6^2 - 
        30 z_3^2 z_6^2 + z_1^2 z_3^2 z_6^2 \cr
	&+ 
        z_1^4 z_3^2 z_6^2 - 18 z_2 z_3^2 z_6^2 - 
        9 z_1^2 z_2 z_3^2 z_6^2 - 
        z_1^4 z_2 z_3^2 z_6^2 - 5 z_2^2 z_3^2 z_6^2 + 
        z_1^2 z_2^2 z_3^2 z_6^2 - 3 z_2^3 z_3^2 z_6^2 - 
        z_1^2 z_2^3 z_3^2 z_6^2 + 28 z_4 z_6^2 - 
        16 z_1^2 z_4 z_6^2 + 2 z_1^4 z_4 z_6^2 + 
        22 z_2 z_4 z_6^2 - 7 z_1^2 z_2 z_4 z_6^2 \cr
	&- 
        z_1^4 z_2 z_4 z_6^2 - 6 z_2^2 z_4 z_6^2 - 
        7 z_2^3 z_4 z_6^2 + z_1^2 z_2^3 z_4 z_6^2 + 
        58 z_3 z_4 z_6^2 - 23 z_1^2 z_3 z_4 z_6^2 + 
        z_1^4 z_3 z_4 z_6^2 + 38 z_2 z_3 z_4 z_6^2 - 
        z_1^2 z_2 z_3 z_4 z_6^2 - 
        z_1^4 z_2 z_3 z_4 z_6^2 - 
        5 z_2^2 z_3 z_4 z_6^2 + 
        z_1^2 z_2^2 z_3 z_4 z_6^2 - 
        3 z_2^3 z_3 z_4 z_6^2\cr
	&-
        z_1^2 z_2^3 z_3 z_4 z_6^2 + 28 z_3^2 z_4 z_6^2 - 
        5 z_1^2 z_3^2 z_4 z_6^2 - 
        z_1^4 z_3^2 z_4 z_6^2 + 18 z_2 z_3^2 z_4 z_6^2 + 
        4 z_1^2 z_2 z_3^2 z_4 z_6^2 + 
        3 z_2^2 z_3^2 z_4 z_6^2 - 
        z_1^2 z_2^2 z_3^2 z_4 z_6^2 + 
        2 z_2^3 z_3^2 z_4 z_6^2 + 2 z_4^2 z_6^2 \cr
	&+ 
        z_1^2 z_4^2 z_6^2 - z_1^4 z_4^2 z_6^2 + 
        2 z_1^2 z_2 z_4^2 z_6^2 + z_2^2 z_4^2 z_6^2 + 
        z_1^2 z_2^2 z_4^2 z_6^2 + 2 z_2^3 z_4^2 z_6^2 - 
        6 z_3 z_4^2 z_6^2 + 5 z_1^2 z_3 z_4^2 z_6^2 - 
        z_1^4 z_3 z_4^2 z_6^2 - 4 z_2 z_3 z_4^2 z_6^2 + 
        2 z_1^2 z_2 z_3 z_4^2 z_6^2 + 
        3 z_2^2 z_3 z_4^2 z_6^2 \cr
	&- 
        z_1^2 z_2^2 z_3 z_4^2 z_6^2 + 
        2 z_2^3 z_3 z_4^2 z_6^2 - 6 z_3^2 z_4^2 z_6^2 + 
        2 z_1^2 z_3^2 z_4^2 z_6^2 - 
        4 z_2 z_3^2 z_4^2 z_6^2 - 6 z_5 z_6^2 + 
        3 z_1^2 z_5 z_6^2 - z_1^4 z_5 z_6^2 + 
        2 z_2 z_5 z_6^2 - 7 z_1^2 z_2 z_5 z_6^2 + 
        z_1^4 z_2 z_5 z_6^2 - z_2^2 z_5 z_6^2 \cr
	&- 
        3 z_1^2 z_2^2 z_5 z_6^2 - 3 z_2^3 z_5 z_6^2 - 
        z_1^2 z_2^3 z_5 z_6^2 + 10 z_3 z_5 z_6^2 - 
        7 z_1^2 z_3 z_5 z_6^2 + z_1^4 z_3 z_5 z_6^2 + 
        6 z_2 z_3 z_5 z_6^2 - z_1^2 z_2 z_3 z_5 z_6^2 - 
        z_1^4 z_2 z_3 z_5 z_6^2 - 
        5 z_2^2 z_3 z_5 z_6^2 + 
        z_1^2 z_2^2 z_3 z_5 z_6^2 - 
        3 z_2^3 z_3 z_5 z_6^2 \cr
	&- 
        z_1^2 z_2^3 z_3 z_5 z_6^2 + 10 z_3^2 z_5 z_6^2 - 
        2 z_1^2 z_3^2 z_5 z_6^2 + 
        6 z_2 z_3^2 z_5 z_6^2 + 
        2 z_1^2 z_2 z_3^2 z_5 z_6^2 + 28 z_4 z_5 z_6^2 - 
        16 z_1^2 z_4 z_5 z_6^2 + 2 z_1^4 z_4 z_5 z_6^2 + 
        6 z_2 z_4 z_5 z_6^2 + z_1^2 z_2 z_4 z_5 z_6^2 - 
        z_1^4 z_2 z_4 z_5 z_6^2 \cr
	&- 
        10 z_2^2 z_4 z_5 z_6^2 - z_2^3 z_4 z_5 z_6^2 - 
        z_1^2 z_2^3 z_4 z_5 z_6^2 + 
        14 z_3 z_4 z_5 z_6^2 + z_1^2 z_3 z_4 z_5 z_6^2 - 
        z_1^4 z_3 z_4 z_5 z_6^2 + 
        8 z_2 z_3 z_4 z_5 z_6^2 + 
        6 z_1^2 z_2 z_3 z_4 z_5 z_6^2 + 
        3 z_2^2 z_3 z_4 z_5 z_6^2 - 
        z_1^2 z_2^2 z_3 z_4 z_5 z_6^2 \cr
	&+ 
        2 z_2^3 z_3 z_4 z_5 z_6^2 - 
        6 z_3^2 z_4 z_5 z_6^2 + 
        2 z_1^2 z_3^2 z_4 z_5 z_6^2 - 
        4 z_2 z_3^2 z_4 z_5 z_6^2 - 14 z_4^2 z_5 z_6^2 + 
        9 z_1^2 z_4^2 z_5 z_6^2 - 
        z_1^4 z_4^2 z_5 z_6^2 - 4 z_2 z_4^2 z_5 z_6^2 + 
        2 z_1^2 z_2 z_4^2 z_5 z_6^2 + 
        7 z_2^2 z_4^2 z_5 z_6^2 \cr
	&- 
        z_1^2 z_2^2 z_4^2 z_5 z_6^2 + 
        2 z_2^3 z_4^2 z_5 z_6^2 - 
        12 z_3 z_4^2 z_5 z_6^2 + 
        4 z_1^2 z_3 z_4^2 z_5 z_6^2 - 
        8 z_2 z_3 z_4^2 z_5 z_6^2 + 14 z_5^2 z_6^2 - 
        5 z_1^2 z_5^2 z_6^2 + 6 z_2 z_5^2 z_6^2 + 
        3 z_1^2 z_2 z_5^2 z_6^2 - z_2^2 z_5^2 z_6^2 - 
        z_2^3 z_5^2 z_6^2 + 16 z_3 z_5^2 z_6^2 \cr
	&+ 
        16 z_2 z_3 z_5^2 z_6^2 + 6 z_3^2 z_5^2 z_6^2 + 
        z_1^2 z_3^2 z_5^2 z_6^2 + 
        6 z_2 z_3^2 z_5^2 z_6^2 + 
        z_1^2 z_2 z_3^2 z_5^2 z_6^2 + 
        z_2^2 z_3^2 z_5^2 z_6^2 + 
        z_2^3 z_3^2 z_5^2 z_6^2 + 4 z_4 z_5^2 z_6^2 + 
        10 z_2 z_4 z_5^2 z_6^2 + 
        z_1^2 z_2 z_4 z_5^2 z_6^2 \cr
	&- 
        2 z_2^2 z_4 z_5^2 z_6^2 - 
        z_2^3 z_4 z_5^2 z_6^2 + 14 z_3 z_4 z_5^2 z_6^2 + 
        z_1^2 z_3 z_4 z_5^2 z_6^2 - 
        2 z_2 z_3 z_4 z_5^2 z_6^2 + 
        z_1^2 z_2 z_3 z_4 z_5^2 z_6^2 + 
        z_2^2 z_3 z_4 z_5^2 z_6^2 + 
        z_2^3 z_3 z_4 z_5^2 z_6^2 + 
        4 z_3^2 z_4 z_5^2 z_6^2 \cr
	&+ 
        z_1^2 z_3^2 z_4 z_5^2 z_6^2 - 
        2 z_2 z_3^2 z_4 z_5^2 z_6^2 + 
        z_2^2 z_3^2 z_4 z_5^2 z_6^2 + 
        6 z_4^2 z_5^2 z_6^2 + z_1^2 z_4^2 z_5^2 z_6^2 - 
        4 z_2 z_4^2 z_5^2 z_6^2 - 
        z_2^2 z_4^2 z_5^2 z_6^2 - 
        2 z_3 z_4^2 z_5^2 z_6^2 + 
        z_1^2 z_3 z_4^2 z_5^2 z_6^2 \cr
	&+ 
        z_2^2 z_3 z_4^2 z_5^2 z_6^2 - 
        2 z_3^2 z_4^2 z_5^2 z_6^2 - 2 z_5^3 z_6^2 + 
        3 z_1^2 z_5^3 z_6^2 + 2 z_2 z_5^3 z_6^2 - 
        z_1^2 z_2 z_5^3 z_6^2 + z_2^2 z_5^3 z_6^2 + 
        z_2^3 z_5^3 z_6^2 - 2 z_3 z_5^3 z_6^2 + 
        z_1^2 z_3 z_5^3 z_6^2 - 2 z_2 z_3 z_5^3 z_6^2 + 
        z_1^2 z_2 z_3 z_5^3 z_6^2 + 
        z_2^2 z_3 z_5^3 z_6^2 + z_2^3 z_3 z_5^3 z_6^2 \cr
	&- 
        2 z_3^2 z_5^3 z_6^2 - 2 z_2 z_3^2 z_5^3 z_6^2 + 
        4 z_4 z_5^3 z_6^2 - 6 z_2 z_4 z_5^3 z_6^2 + 
        z_1^2 z_2 z_4 z_5^3 z_6^2 + 
        2 z_2^2 z_4 z_5^3 z_6^2 + 
        z_2^3 z_4 z_5^3 z_6^2 - 6 z_3 z_4 z_5^3 z_6^2 + 
        z_1^2 z_3 z_4 z_5^3 z_6^2 - 
        4 z_2 z_3 z_4 z_5^3 z_6^2 + 
        z_2^2 z_3 z_4 z_5^3 z_6^2 \cr
	&- 
        2 z_3^2 z_4 z_5^3 z_6^2 - 10 z_4^2 z_5^3 z_6^2 + 
        z_1^2 z_4^2 z_5^3 z_6^2 + 
        z_2^2 z_4^2 z_5^3 z_6^2 - 
        4 z_3 z_4^2 z_5^3 z_6^2 + 5 z_6^4 - z_1^2 z_6^4 + 
        3 z_2 z_6^4 + z_1^2 z_2 z_6^4 + 10 z_3 z_6^4 - 
        2 z_1^2 z_3 z_6^4 + 6 z_2 z_3 z_6^4 + 
        2 z_1^2 z_2 z_3 z_6^4 + 5 z_3^2 z_6^4 - 
        z_1^2 z_3^2 z_6^4 + 3 z_2 z_3^2 z_6^4 \cr
	&+ 
        z_1^2 z_2 z_3^2 z_6^4 + 2 z_4 z_6^4 + 
        z_2 z_4 z_6^4 + z_1^2 z_2 z_4 z_6^4 - 
        z_3 z_4 z_6^4 + z_1^2 z_3 z_4 z_6^4 - 
        z_2 z_3 z_4 z_6^4 + z_1^2 z_2 z_3 z_4 z_6^4 - 
        3 z_3^2 z_4 z_6^4 + z_1^2 z_3^2 z_4 z_6^4 - 
        2 z_2 z_3^2 z_4 z_6^4 - 3 z_4^2 z_6^4 + 
        z_1^2 z_4^2 z_6^4 - 2 z_2 z_4^2 z_6^4 - 
        3 z_3 z_4^2 z_6^4 + z_1^2 z_3 z_4^2 z_6^4 \cr
	&- 
        2 z_2 z_3 z_4^2 z_6^4 + 5 z_5 z_6^4 - 
        z_1^2 z_5 z_6^4 + 3 z_2 z_5 z_6^4 + 
        z_1^2 z_2 z_5 z_6^4 + 5 z_3 z_5 z_6^4 - 
        z_1^2 z_3 z_5 z_6^4 + 3 z_2 z_3 z_5 z_6^4 + 
        z_1^2 z_2 z_3 z_5 z_6^4 + 2 z_4 z_5 z_6^4 + 
        z_2 z_4 z_5 z_6^4 + z_1^2 z_2 z_4 z_5 z_6^4 - 
        3 z_3 z_4 z_5 z_6^4 + z_1^2 z_3 z_4 z_5 z_6^4 - 
        2 z_2 z_3 z_4 z_5 z_6^4 - 3 z_4^2 z_5 z_6^4 \cr
	&+ 
        z_1^2 z_4^2 z_5 z_6^4 - 2 z_2 z_4^2 z_5 z_6^4 - 
        z_5^2 z_6^4 - z_2 z_5^2 z_6^4 - 
        2 z_3 z_5^2 z_6^4 - 2 z_2 z_3 z_5^2 z_6^4 - 
        z_3^2 z_5^2 z_6^4 - z_2 z_3^2 z_5^2 z_6^4 - 
        2 z_4 z_5^2 z_6^4 - z_2 z_4 z_5^2 z_6^4 - 
        3 z_3 z_4 z_5^2 z_6^4 - 
        z_2 z_3 z_4 z_5^2 z_6^4 - 
        z_3^2 z_4 z_5^2 z_6^4 - z_4^2 z_5^2 z_6^4 \cr
	&- 
        z_3 z_4^2 z_5^2 z_6^4 - z_5^3 z_6^4 - 
        z_2 z_5^3 z_6^4 - z_3 z_5^3 z_6^4 - 
        z_2 z_3 z_5^3 z_6^4 - 2 z_4 z_5^3 z_6^4 - 
        z_2 z_4 z_5^3 z_6^4 - z_3 z_4 z_5^3 z_6^4 - 
        z_4^2 z_5^3 z_6^4)\cr}}

\end